\documentclass[journal]{IEEEtran}
\usepackage{graphicx,amsmath,amssymb}
\usepackage{subfigure}
\usepackage{citesort}
\usepackage{fancyhdr}
\usepackage{mdwmath}
\usepackage{mdwtab}
\usepackage{balance}
\usepackage{xcolor}
\usepackage{bm}
\usepackage{amsthm}
\usepackage{algorithm}
\usepackage{algorithmic}
\usepackage{multirow}
\usepackage{flafter}
\usepackage{cite}

\newtheorem{remark}{Remark}
\newtheorem{proposition}{Proposition}
\newtheorem{theorem}{Theorem}

\newtheorem{lemma}{Lemma}

\newtheorem{corollary}{Corollary}

\newtheorem{assumption}{Assumption}

\begin{document}
\title{A Unified Spatial Framework for UAV-aided MmWave Networks}

\author{Wenqiang~Yi,~\IEEEmembership{Student Member,~IEEE,}
        Yuanwei~Liu,~\IEEEmembership{Senior Member,~IEEE,}
        Eliane~Bodanese,~\IEEEmembership{Member,~IEEE,}
        Arumugam~Nallanathan,~\IEEEmembership{Fellow,~IEEE,}
        and George~K.~Karagiannidis,~\IEEEmembership{Fellow,~IEEE}
\thanks{W. Yi, Y. Liu, E. Bodanese and A. Nallanathan are with the Queen Mary University of London, London E1 4NS, UK (email:\{w.yi, yuanwei.liu, eliane.bodanese, a.nallanathan\}@qmul.ac.uk).}
\thanks{G. K. Karagiannidis is with the Aristotle University of Thessaloniki, Thessaloniki 541 24, Greece (email: geokarag@auth.gr).}
\thanks{Part of this work was presented in IEEE Global Communications Conference (GLOBECOM), December, UAE, 2018~\cite{123456789}.}}
\maketitle

\begin{abstract}
  For unmanned aerial vehicle (UAV) aided millimeter wave (mmWave) networks, we propose a unified three-dimensional (3D) spatial framework in this paper to model a general case that uncovered users send messages to base stations via UAVs. More specifically, the locations of transceivers in downlink and uplink are modeled through the Poisson point processes and Poisson cluster processes (PCPs), respectively. For PCPs, Matern cluster and Thomas cluster processes, are analyzed. Furthermore, both 3D blockage processes and 3D antenna patterns are introduced for appraising the effect of altitudes. Based on this unified framework, several closed-form expressions for the coverage probability in the uplink and downlink, are derived. By investigating the entire communication process, which includes the two aforementioned phases and the cooperative transmission between them, tractable expressions of system coverage probabilities are derived. Next, three practical applications in UAV networks are provided as case studies of the proposed framework. The results reveal that the impact of thermal noise and non-line-of-sight mmWave transmissions is negligible. In the considered networks, mmWave outperforms sub-6~GHz in terms of the data rate, due to the sharp direction beamforming and large transmit bandwidth. Additionally, there exists an optimal altitude of UAVs, which maximizes the system coverage probability.
\end{abstract}


\section{Introduction}

 Due to the stationary locations and high cost of traditional macro base stations (BSs), it is extremely arduous to provide ubiquitous coverage via terrestrial cellular networks, especially for critical applications, e.g. disaster rescue, firefighting, reconnaissance, etc~\cite{7470933,scenarios,7470937}. Therefore, an effective solution is urgently needed for the next generation of wireless networks. Under these circumstances, unmanned aerial vehicles (UAVs) become increasingly popular, owing to their flexibility and autonomy~\cite{7995044}. In terms of the frequency band used for UAV networks, two main candidates have been proposed: sub-6 GHz and millimeter wave (mmWave)~\cite{7470933,7470937,8255824}. Benefited by mature wireless techniques and tolerance to blockages, the existing sub-6 GHz in traditional networks can be reused effortlessly in a few dense-obstacle environments~\cite{7470933}. However, in most of the cases, UAVs are able to establish line-of-sight (LOS) connections to users by adjusting their locations and altitudes~\cite{7470937}. Under this condition, mmWave bands become the best choice for UAV-aided networks due to its larger available bandwidth and consequent higher data rate transmission in LOS  than sub-6 GHz~\cite{8401954}. Moreover, with the aid of large antenna scales deployed at mmWave devices, sharp directional beams can be generated to increase network capacity~\cite{6932503} and mitigate the Doppler spread~\cite{7470937}.

 In order to analyze the average performance of wireless networks, stochastic geometry has rekindled the strong interest of academia~\cite{7982794}. More specifically, the locations of transceivers are distributed according to different random spatial point processes, which are capable of statistically depicting the nature of wireless systems~\cite{7445146}. One popular point pattern is Poisson point process~(PPP), which has been widely adopted in recent works for modeling network nodes such as BSs and multiple pieces of user equipment~\cite{7445146,maamari2016coverage}. The main advantage of this pattern is that all nodes are uniformly distributed in a plane. However, in practical scenarios, most wireless systems have a clustered property. For example, in device-to-device (D2D) communications or cognitive networks, users with the same demands are frequently gathered to form a cluster and the PPP model fails to accurately describe this property. As a result, another point pattern named Poisson cluster process (PCP), which includes multiple clusters, becomes increasingly attractive in the literature~\cite{5208529,8016632,8635489}.

\subsection{State-of-the-Art and Motivation}

The performance evaluation of UAV networks plays a vital role in proposing new relative protocols and designing efficient network structures. Accordingly, Al-Hourani \emph{et~al.}~\cite{6863654} provided a theoretical approach to obtain the optimal altitude of UAVs, with an aim for maximizing the coverage on the ground. To appraise the path losses and antenna effects caused by the randomness of UAV networks, Mozaffari \emph{et~al.}~\cite{7412759} investigated the average coverage probability and throughput in a UAV-aided network with underlaid D2D communications, where the D2D users were modeled as a homogeneous PPP and the downlink users were uniformly distributed in a finite area forming a variant of PPP, known as binary Poisson process (BPP). Chetlur and Dhillon~\cite{7967745} extended this work to a multi-UAV case, where the locations of UAVs obey a BPP. With the aid of this framework, the coverage performance of downlink transmissions was analysed. Recently, the authors in~\cite{7756327} introduced a three-dimensional~(3D) PPP incorporating the adjustable altitude, in order to model the locations of UAVs. Analytical expressions for the coverage probability of drone small-cells were derived. For disaster scenarios that requiring extending the coverage of UAVs, Zhao \emph{et~al.}~\cite{8641424} proposed a novel framework for evaluating UAV-assisted emergency networks with multihop D2D communications. Regarding multiple access techniques, the performance of UAV networks with non-orthogonal multiple access (NOMA) techniques was studied in~\cite{8629316,8488592}, which can be used in applications with massive users. Due to the limited onboard energy, optimizing power allocation strategies becomes important. The optimal strategies have been provided in various UAV applications, such as securing UAV communications~\cite{8618602,8392472}, multi-hop UAV relaying communications~\cite{8453022}, and so forth. In addition, promising machine learning approaches have been adopted to design the movement and power control in multi-UAV assisted networks~\cite{8727504,8736350}.

 In terms of the frequency band utilized for UAV networks, the aforementioned research contributions~\cite{6863654,7412759,7967745,7756327} focused on the sub-6 GHz scenarios. However, when considering mmWave scenarios, blockage environment, antenna patterns, and fading channel model should be carefully modified. Instead of using the ray tracing method as discussed in~\cite{8268122}, Bai \emph{et al.}~\cite{6840343} leveraged the random shape theory to propose a mathematical blockage model for high frequencies. This model was approximated by a fixed line-of-sight (LOS) disc in~\cite{6932503}, which had acceptable accuracy with high calculation efficiency. By invoking the altitude information of devices, the conventional two-dimensional (2D) blockage model was extended to the 3D case with the aid of Rayleigh distributed buildings~\cite{LOSmodel}. In addition to the widely used omnidirectional antenna pattern in sub-6~GHz, Zhu \emph{et al.}~\cite{8335329} introduced a 3D sectorized beamforming pattern for mmWave antennas to depict the sharp directional beam. Furthermore, due to the huge difference between LOS and non-LOS (NLOS) mmWave transmissions, the Nakagami-$m$ fading channel was preferable in mmWave-enabled networks. This small-scale fading model was considered in~\cite{maamari2016coverage,7931577,8335329,6932503,8016632}.

 In UAV-aided networks, for the communications between terrestrial BSs and UAVs, the locations of transceivers can be modeled as two independent PPPs to capture the randomness of the networks. However, for communications between UAVs and users, in most of the cases one UAV is dedicated to serve a cluster of users with the same requirements. Recently, Haenggi~\cite{7852435} offered a soft-core process by using the usual K function by Ripley to model uplink users, but the exact distribution of communication distances is not efficient when calculating coverage probabilities. Therefore, we utilize another popular pattern, namely PCP, to model these scenarios. Furthermore, Ganti and Haenggi~\cite{5208529} first proposed PCP to model the node locations in clustered wireless ad hoc networks. In this work, two specific models, i.e., \emph{Matern cluster processes (MCPs)} and \emph{Thomas cluster processes (TCPs)}, were provided. Then, several networks were studied with these two types of PCPs, e.g. cognitive networks~\cite{6155562}, heterogeneous networks (HetNets)~\cite{7110502}, and D2D networks~\cite{8016632}. To the best of our knowledge, the research in the context of UAV networks by utilizing PCP is still in its infancy.

 In the real world, the distribution of users are decided by the users' purposes and hence it is usually independent on the distribution of BSs. When considering the mobility of UAVs and the randomness of user clusters, it is challenging to create a general spatial framework including both uplink and downlink phases for large-scale UAV-aided networks. Additionally, the evaluation of average coverage performance in UAV networks, especially for mmWave scenarios, is still at the very early stage. These two factors are key motivations for this article.

\subsection{Contributions and Organization}

 We propose a unified spatial framework for UAV-aided networks, where a UAV first collects messages from one user in an \emph{uplink phase} and then it retransmits the desired information to the requiring BS in a \emph{downlink phase}. The transceivers in the downlink phase and uplink phase are distributed with the aid of a PPP and a PCP, respectively. Particularly, two commonly used PCPs, namely TCP and MCP, are investigated, and mmWave communications are considered as well. In summary, the main contributions of this paper are as follow:
 \begin{itemize}
   \item We propose an analytical framework for UAV-aided networks with multiple clustered users. Dissimilar to that for traditional ground networks, a 3D blockage process and a 3D up-tilted antenna model are introduced to characterize the impact of UAVs' altitudes and positions. Additionally, both uplink and downlink phases are assessed in order to model the entire transmission process of UAV-aided communications.
   \item We investigate the distributions of communication distances in the downlink phase. Based on these distributions, closed-form expressions for the coverage probability are derived to enhance the evaluation efficiency. Regarding the uplink phase, we derive a general expression for coverage probabilities. Moreover, this expression is able to characterize TCP as well as MCP. Note that the number of users in each cluster can be a constant or to follow a Poisson distribution depending on different network requirements. We theoretically demonstrate that these deployment scenarios perform similarly when the number of interferers is large.
   \item We provide a practical cooperation transmission to model the behaviours of UAVs from the uplink phase to the downlink phase. Three typical applications of UAVs are examined with the aid of the unified framework, and thus validating the flexibility of our work. In order to study the performance of the entire communication process, we also derive tractable expressions for the system coverage probability.
   \item We show that, 1) when the geographical size of clusters and the interfering number of users are small, the inter-cluster interference can be ignored to simplify the analysis; 2) the effect of thermal noise is negligible in the proposed UAV networks. In terms of mmWave communications, NLOS transmissions can be ignored as well, especially in a low-density blockage environment; 3) there exists an optimal altitude of UAVs for achieving the maximum system coverage probability; and 4) a large antenna scale is able to enlarge its main beam gain and narrow the beamwidth for compensating path loss, thereby enhancing coverage performance.
 \end{itemize}

The rest of this paper is organized as follows: Section~II proposes the network model including the PPP distributed downlink phase and the PCP distributed uplink phase. Section~III first describes the coverage performance in the proposed UAV-aided networks. Then, three practical applications are studied with the aid of the unified framework. Section~IV provides the validating and numerical results. Section~V presents the conclusions.

\section{Network Model}

We introduce a \emph{unified spatial framework} for UAV-aided networks by integrating mmWave communications, which describes a general case where BSs are located far from the served users, and hence UAVs are deployed to establish the desired communication links. Due to the mobility of UAVs, we assume that multiple serving UAVs are able to fly from the user region to the BS region. To evaluate the average performance of the proposed networks, one reference transmission from a \emph{transmitting user} to a \emph{typical BS} via a \emph{corresponding UAV} is randomly selected from all possible links\footnote{In this paper, we focus on the coverage probability which is an important metric to evaluate system throughput~\cite{8401954}, outage probabilities~\cite{7445146}, area spectrum efficiency~\cite{8016632}, etc. The analysis of delay will be included in our future work.}. The entire transmission process can be divided into two phases: 1) \emph{Uplink Phase}, where the transmitting user uploads the desired message to the corresponding UAV; and 2) \emph{Downlink Phase}, where the typical BS downloads the required message from the corresponding UAV. These two phases are processed in different time slots. Regarding the flight process, three practical cooperation transmission scenarios are discussed at the end of the next section.

\subsection{Spatial Distribution}

 \begin{figure*}[t!]
  \centering
  \includegraphics[height= 2 in]{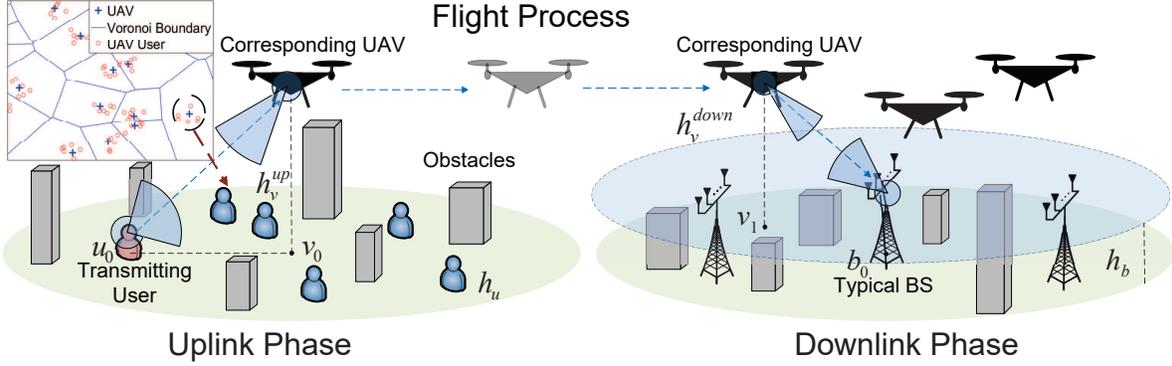}\\
  \caption{The layout of proposed UAV-aided networks with the uplink and downlink phases. }\label{fig1}
 \end{figure*}
 Different spatial distributions are assumed for the two phases. For ease of understanding, we first introduce the downlink and then the uplink phase.
\subsubsection{Downlink Phase}
 In the downlink phase, since BSs are able to provide ubiquitous coverage, we consider a general scenario, where UAVs control themselves without extra control and location information from BSs and hence their trajectories are independent of the distribution of BSs. Therefore, we assume that multiple macro BSs and UAVs are modeled as two independent PPPs with density ${\lambda_b}$ and ${\lambda^{down}_v}$, denoted by ${\Phi_b}$ and ${\Phi^{down}_v}$, respectively.\footnote{The considered scenario can be simply extended to other BS controlling cases by introducing a controlling parameter $C_{\rm {BS}}$ to generate a thinning process of $\Phi_v^{down}$ with a density $C_{\rm {BS}}\lambda^{down}_v$.}. Note that BSs also need to serve nearby ground users in practise. We assume BSs communicate with the nearby users via different resource blocks to avoid extra interference. Regarding the altitude of each device, UAVs are hovering at a height of $h^{down}_v$, while the receiving antennas of macro BSs are located at an altitude $h_b$, as illustrated in Fig.~\ref{fig1}.
\subsubsection{Uplink Phase}
 For the uplink phase, since users with similar requests frequently stay together, we utilize PCPs to model this cluster property~\cite{5208529}. In one PCP, \emph{parent points} are distributed following a homogeneous PPP ${\Phi^{up}_v}$ with density ${\lambda^{up}_v}$. Around each parent point at $v\in \Phi^{up}_v$ $(v \in \mathbb{R}^2)$, \emph{daughter points} (denoted by $\mathbb{U}_v$) are independently and identically distributed~(i.i.d.) forming a cluster. The number of the daughter points can be constant, $\bar N$, or a Poisson random variable with mean $\bar n$ for different purposes. Due to the mobility and limited onboard energy, UAVs are more suitable for serving hot spots than providing ubiquitous coverage. When users with similar interests are grouped into one cluster, it is reasonable to locate a UAV in the center to serve several intra-cluster users simultaneously. Therefore, we assume daughter points represent locations of users. Moreover, UAVs are located at parent points to provide a central controlling structure.

 Note that TCP and MCP are two widely used patterns of PCPs. The main difference between them is the distribution of daughter points in each cluster. For TCP, these points are modeled as a symmetrical normal distribution around the central parent point, with a standard deviation $\sigma$. For MCP, the daughter points are uniformly distributed in a disc with radius $R$. The two density functions for the distance between a user at $u\in \mathbb{U}_v$ $(u \in \mathbb{R}^2)$ to the above UAV at $v$ can be expressed as
\begin{align}\label{1}
&{f^{Tho}_U}\left( u \right) = \frac{1}{{2\pi {\sigma ^2}}}\exp \left( { - \frac{{{{\left\| u-v \right\|}^2}}}{{2{\sigma ^2}}}} \right),\\
&f_U^{Mat}\left( u \right) = \frac{1}{{\pi {R^2}}}\mathbf{U}\left( {R - \left\| {u - v} \right\|} \right),
\end{align}
where the superscript $Tho$ represents TCP and $Mat$ represents MCP. The $\mathbf{U}(.)$ is the unit step function, which is given by
\begin{align}
\mathbf{U}\left( x \right) = \left\{ {\begin{array}{*{20}{c}}
{1,}&{x \ge 0}\\
{0,}&{x < 0}
\end{array}} \right..
\end{align}

Regarding the altitude in the uplink phase, we assume that all users are located at the ground with an altitude $h_u$ and each UAV hovers above one cluster center at an altitude $h^{up}_v$ to serve the intra-cluster users.

\subsection{Blockage Model}

The main difference between sub-6 GHz and mmWave channels is the blockage effect~\cite{6932503}. Since mmWave signals are more sensitive to obstacles than sub-6 GHz signals, the blockage model is important for mmWave communications. The authors in~\cite{LOSmodel} have proposed a tractable blockage model. The extracted theoretical expressions are derived based on a three dimensional obstacle environment, where the average height and the density of obstacles are adjustable. Therefore, this model is suitable for various practical scenarios, e.g., the suburban with low-density and low-altitude obstacles, the urban with dense and high buildings, and so forth. In this model, the density of obstacles and the ratio of the obstacle area to the total area are represented by $\beta_b$ m$^{-2}$ and $\beta_a$, respectively. The height of each obstacle is modeled as a Rayleigh distribution with a scale parameter~$\varepsilon$. When the horizontal transmission distance is $r$, then the probability density function (PDF) of a LOS link is given by~\cite{stochastic}
\begin{align}\label{LOSp}
&{p_L}\left( {\gamma (r)\left| {{h_t},{h_r}} \right.} \right) = \prod\limits_{n = 0}^{\max \left( {0,\gamma(r) } \right)} {\bigg( {1 - }}\nonumber \\
&{{\exp \bigg( { - \frac{{{{\left( {\gamma(r) \max \left( {{h_t},{h_r}} \right) - \left( {n + {1 \mathord{\left/
 {\vphantom {1 2}} \right.
 \kern-\nulldelimiterspace} 2}} \right)\left| {{h_t} - {h_r}} \right|} \right)}^2}}}{{2{\varepsilon ^2}{\gamma(r) ^2}}}} \bigg)} \bigg),}
\end{align}
where $\gamma (r)  = \left\lfloor {r\sqrt {{\beta_a}{\beta_b}}} \right\rfloor$ and $\lfloor.\rfloor$ is the floor function. As $\gamma(r)$ is not a continuous function in terms of $r$ but nonnegative integers, namely $\gamma(r)\in \mathbb{Z}^*$ ($\mathbb{Z}^*$ is the set of nonnegative integers), we use $\gamma$ instead of $\gamma(r)$ in the rest of this paper. The subscript $t$ and $r$ represent transmitters and receivers, respectively\footnote{In the downlink phase, the transmitters are UAVs and the receivers are BSs, while in the uplink phase, $t$ and $r$ represent users and UAVs, respectively. We omit this explanation in the rest of this paper.}. Intuitively, the PDF of a NLOS link is $p_N(.)=(1-{p_L}(.))$. Based on several propagation measurements for mmWave communications~\cite{6655399,rappaport201238}, the path loss law in the UAV network can be expressed as
\begin{align}\label{LOSorNLOS}
&\mathrm{L}(r\left| {{h_t},{h_r}} \right.) \nonumber \\
= &\frac{{\mathbb{B}\left( {{p_L}\left( {\gamma \left| {{h_t},{h_r}} \right.} \right)} \right){C_L}}}{{\sqrt {{{\left( {{r^{\rm{2}}} + |{h_t} - {h_r}{|^2}} \right)}^{{\alpha _L}}}} }} + \frac{{ { \mathbb{B}\left( {{p_N}\left( {\gamma \left| {{h_t},{h_r}} \right.} \right)} \right)} {C_N}}}{{\sqrt {{{\left( {{r^{\rm{2}}} + |{h_t} - {h_r}{|^2}} \right)}^{{\alpha _N}}}} }},
\end{align}
where $\mathbb{B}(x)$ is a Bernoulli random variable with parameter of success probability $x$. The parameters $C_\kappa$ and $\alpha_\kappa$ are the considered intercept and path loss exponent, respectively. Additionally, $\kappa \in \{L, N\}$, where $L$ represents LOS links and $N$ denotes NLOS links.

\subsection{Directional Beamforming}

\begin{figure}[t!]
  \centering
  \includegraphics[height= 1.5 in]{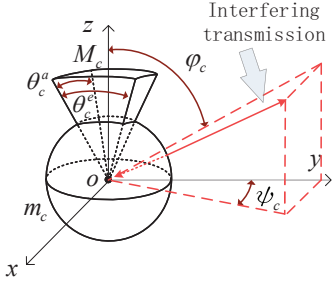}\\
  \caption{The sectorized antenna pattern. }\label{fig3D}
 \end{figure}

We consider the typical sectorized antenna pattern, as mentioned in~\cite{7491363} at all transceivers to accomplish the sharp directional beamforming (see Fig.~\ref{fig3D}). In order to guarantee that UAVs can be benefited from the high main-beam gain, we assume BSs utilized up-tilted antennas to serve UAVs. With the aid of stabilizers and machine-leaning-based beam tracking techniques~\cite{8662770}, the misalignment caused by vibration can be mitigated. Moreover, since we evaluate the average antenna beam gain in our systems, the effect of misalignment is ignored in the rest of this paper. The numbers of antenna elements at macro BSs, UAVs and users are $N_b$, $N_v$ and $N_u$, respectively. Four main characteristics of antennas are introduced in this pattern, which are the half-power beamwidth in the azimuth plane $\theta_c^a$, the counterpart in the elevation plane $\theta_c^e$, the main beam gain $M_c$, the side lobe gain $m_c$  ($c\in\{t,r\}$). In each resource block, we assume every receiver serves one transmitter at once. Based on the hierarchical beam search scheme~\cite{5262295}, the location information of devices can be obtained at both the transmitter and the receiver. Then, they adjusts the antenna directions toward each other for achieving the maximum beamforming gain $G_0$. When the receiver turns to another transmitter, both of them need to change their antenna directions to obtain $G_0$. It is worth noting that $G_0$ has a positive correlation with the antenna scale at the considered transmitter and receiver and hence $G_0=M_tM_r$.

Regarding one interfering transmission, the angles deviating from the boresight direction in the azimuth plane and the elevation plane are $\psi_c$ and $\varphi_c$, respectively. In various wireless systems, e.g. HetNets, D2D communications, and so forth, the interferers may communicate with other devices (macro BSs, pico BSs or users). To enhance the generality, we assume that $\psi_c$ is uniformly distributed in the range $[0,2\pi)$ and $\varphi_c$ is uniformly distributed in the range $[0,\pi]$~\cite{8335329}. The directivity gain at one receiver with the interfering transmitter located at $l$ can be expressed as follows:
\begin{align}
{G_l} = {G({{\theta _t^a},{\theta_t^e},{M_t},{m_t}}}){G({{\theta _r^a},{\theta_r^e},{M_r},{m_r}}}),
\end{align}
where  $ {G({{\theta _c^a},{\theta_c^e},{M_c},{m_c}}})$ denotes the directional antenna gain.  Therefore, ${G_l}$ has four patterns as shown in Table.~\ref{tab1}. Each pattern has the value $o_i$ with the probability $p_i$, where $i \in \left\{ {1,2,3,4} \right\}$ and $G_0=o_1$.

\begin{table*}[ht!]
\centering
\caption{Four patterns of $G_l$}
\label{tab1}
\begin{tabular}{c|c|c|c|c}
\hline
\hline
  $i$     &$1$        &$2$        &$3$        &$4$\\ \hline
  $o_i$ &$M_tM_r$ &$m_tM_r$ &$M_tm_r$ &$m_tm_r$\\  \hline
  $p_i$ &$\big( {\frac{{\theta _t^a}}{{2\pi }}  \frac{{\theta _t^e}}{\pi }} \big)  \big( {\frac{{\theta _r^a}}{{2\pi }}  \frac{{\theta _r^e}}{\pi }} \big)$ &$\big( {1 - \frac{{\theta _t^a}}{{2\pi }} \frac{{\theta _t^e}}{\pi }} \big)  \big( {\frac{{\theta _r^a}}{{2\pi }} \frac{{\theta _r^e}}{\pi }} \big)$ &$\big( {\frac{{\theta _t^a}}{{2\pi }}  \frac{{\theta _t^e}}{\pi }} \big)  \big( {1 - \frac{{\theta _r^a}}{{2\pi }}  \frac{{\theta _r^e}}{\pi }} \big)$ &$\big( {1 - \frac{{\theta _t^a}}{{2\pi }}  \frac{{\theta _t^e}}{\pi }} \big)  \big( {1 - \frac{{\theta _r^a}}{{2\pi }} \frac{{\theta _r^e}}{\pi }} \big)$\\ \hline
 \hline
\end{tabular}
\end{table*}

\subsection{Signal Model}

In this paper, we consider a general air-ground channel model for both UAV-BS and UAV-user scenarios, which includes antenna gain, small-scale fading, and path loss. Based on the previous discussion, when a receiver is located at the origin with a height $h_r$, the channel gain for a transmitter at $x_t \in \mathbb{R}^2$ with a height $h_t$ is
\begin{align}
H_g({x_r},h_r, h_t) = \mathrm{L}(\|x_r\|\left|h_r, h_t\right.)G_{x_r}|\hat h_{x_r}|^2,
\end{align}
where $\hat h$ is the small scale gain for Nakagami fading channel with parameter $\mathcal{N}_\kappa$ such that $|\hat h|^2$ is a normalized Gamma variable. When the channel is the desired channel, namely the transmission has the maximum antenna gain, the channel gain is changed to $H_g^0({x_r},h_r, h_t) = \mathrm{L}(\|x_r\|\left|h_r, h_t\right.)G_0|\hat h_{x_r}|^2$.

\subsubsection{Downlink Phase}
We assume that the number of BSs is greater than that of UAVs, namely $\lambda_b>\lambda_v^{down}$, such that all UAVs in the downlink phase are active at the considered time slot. The typical BS at $b_0$ $(b_0 \in \Phi_b, b_0 \in \mathbb{R}^2)$ is randomly selected and it is fixed at the origin of the downlink plane. To achieve the best quality of service, the corresponding UAV, which is located at $v_1$ $(v_1 \in \Phi_v^{down}, v_1 \in \mathbb{R}^2)$, is the closest UAV to the typical BS.  Since the interfering signals are offered by the remaining UAVs, the received SINR at the typical BS can be expressed as
\begin{align}\label{downlinkSINR}
{\Upsilon _{down} = \frac{H_g^0({v_1},h_v^{down},h_b)}{{\sum\limits_{v \in \Phi _v^{down}\backslash{v_1}} H_g({v},h_v^{down},h_b)  + n_0^2/P_v}},}
\end{align}
where $P_v$ is the transmit power for each UAV. The power of thermal noise obeys $n_0^2= k_b T_r B$, where $k_b$ is Boltzmann's constant, $T_r$ is the absolute temperature of resistors, and $B$ is the considered bandwidth. In this paper, we assume $T_r = 300$ K so that $n_0^2\approx 4.14\times 10^{-21}B$ W.
\subsubsection{Uplink Phase}
 Since the downlink and uplink phases are analyzed in two independent planes, the corresponding UAV can belong to any of the clusters in the uplink plane. We assume that this UAV is located at the origin of the uplink plane and its location is denoted by $v_0$ $(v_0 \in \Phi_v^{up})$. In one time slot, we randomly select an intra-cluster user at $u_0$ $(u_0 \in \mathbb{U}_{v_0})$ to be the transmitting user. The random selection scheme aims to provide a fair law, where every intra-cluster user has the same opportunity to be served~\cite{7445146}. Moreover, all users are active to represent a full load case. In contrast to the downlink phase, the interference in the uplink phase is originated by both intra-cluster and inter-cluster users. Then, the received SINR at the corresponding UAV is given at the top of next page.
  \begin{figure*}[ht!]
\normalsize
\begin{align}\label{SINRup}
{\Upsilon_{up}} = \frac{H_g^0({u_0},h_u,h_u^{up})}{{\sum\limits_{u \in {\mathbb{U}_{{v_0}}}\backslash{u_0}} {H_g({u},h_u,h_v^{up})}  + \sum\limits_{v \in \Phi _v^{up}\backslash{{v}_0}} {\sum\limits_{u \in {\mathbb{U}_v}} {H_g({u},h_u,h_v^{up})} }  + n_0^2/P_u}}.
\end{align}
\hrulefill \vspace*{0pt}
\end{figure*}
In \eqref{SINRup}, $P_u$ is the transmit power for each user.

\section{Performance Evaluation for UAV-aided Networks}

 In this section, we first evaluate the performance of UAV-aided networks with the aid of the proposed unified framework. Then, we provide several practical scenarios which can be modeled by the proposed framework after minor adjustments.

\subsection{Performance of the Downlink Phase}

 In this part, we analyze the coverage performance of downlink transmissions based on the distribution of distances\footnote{In the remainder of this paper, unless other specified, the distance stands for the horizontal transmission distance.} and Laplace transform of interference.
\subsubsection{Distance Distributions in Poisson Point Processes}
 Assuming that the distance realizations of UAVs in the downlink phase form a set $r_n=\left\|v_n\right\|_{n=1,2,...,N_{dl}}$, where $v_n \in \Phi_v^{down}$ and $N_{dl}$ is the number of UAVs. The subscript $n$ is the parameter of ranked distances, namely $r_1<r_2<...<r_{N_{dl}}$. Note that the corresponding UAV is the closest node to the reference BS. Therefore, the PDF of the nearest communication distance $r_1$ is given by~\cite{7445146}
\begin{align}\label{PDFPPP}
{f_n}\left( {{r_1}} \right) = 2\pi \lambda _v^{down}{r_1}\exp \left( { - \pi \lambda _v^{down}r_1^2} \right).
\end{align}

 In addition to the corresponding UAV, the rest of UAVs which are located further than $r_1$ are interfering transmitters. Due to the Slivnyak's theorem, the density of interfering UAVs is still $\lambda_v^{down}$ in the area $\mathbb{O}(r_1,+\infty)$, where $\mathbb{O}(a,b)$ represents an annulus with the inner radius $a$ and outer radius $b$~\cite{baccelli2009stochastic}. Moreover, each interferer is distributed independently such that the index $n$ can be dropped from $r_n$. Therefore, the PDF of the interfering distance $r$ can be derived via the probability generating function of PPP~\cite{stoyanstochastic}.

\subsubsection{Laplace Transform of Interference in Poisson Point Processes}
For simplifying the notation, we define the Laplace transform of interference as follows:
\begin{align}\label{laplace}
\mathcal{L}_{down}(s)=\mathbb{E}\left[\exp(-sI)\right],
\end{align}
where $\mathbb{E}\left[.\right]$ is the expectation function, $s$ is the transform parameter, and $I$ represents the received power of interference, which is given by
\begin{align}
{I} = \sum\limits_{v \in \Phi _v^{down}\backslash{v_1}} {H_g(v,h_v^{down},h_b)}.
\end{align}
Next, we provide closed-form expressions for the Laplace transform of interference in the following lemma and corollary.
\begin{lemma}\label{lemma1}
\emph{Since the corresponding UAV at $v_1$ is the closest node to the typical BS with a communication distance $r_1=\left\|v_1\right\|$, all interfering UAVs are located in the area $\mathbb{O}(r_1,+\infty)$, namely $r>r_1$. A tractable expression for the Laplace transform of interference in the downlink phase conditional on $r_1$ is given by}
\begin{align}\label{mmWaveLaplace}
&\mathcal{L}_{down} \left( {s\left| {{r_1}} \right.} \right)\nonumber \\
=&\exp \Big( { - 2\pi \lambda _v^{down}\sum\limits_{i = 1}^4 {{p_i}} \left( {\Theta^L\left( {s,{o_i}\left| {{r_1}} \right.} \right) + \Theta^N\left( {s,{o_i}\left| {{r_1}} \right.} \right)} \right)} \Big),
\end{align}
\emph{where}
\begin{align}
&\Theta^\kappa \left( {s,{o_i}\left| {{r_1}} \right.} \right)= {p_\kappa }\left( {\gamma \left| {h_v^{down},{h_b}} \right.} \right){{\rm{Z}}^\kappa }\left( {s,{r_1},\frac{{\gamma  + 1}}{{\sqrt {{\beta _a}{\beta _b}} }},{o_i}} \right) \nonumber \\
&  + \sum\limits_{j = \gamma  + 1}^\infty  {{p_\kappa }\left( {j\left| {h_v^{down},{h_b}} \right.} \right){{\rm{Z}}^\kappa }\left( {s,\frac{j}{{\sqrt {{\lambda _a}{\lambda _b}} }},\frac{{j + 1}}{{\sqrt {{\beta _a}{\beta _b}} }},{o_i}} \right)}
\end{align}
\emph{and}
\begin{align}
{{\rm{Z}}^\kappa }( {s,a,b,{G_v}} ) = &\frac{{{a^2} + \Delta h_d^2}}{2}F_{{\alpha _\kappa }}^\kappa \Big( {\frac{{s{C_\kappa }{G_v}}}{{{\mathcal{N}_\kappa }{{( {{a^2} + \Delta h_d^2} )}^{\frac{{{\alpha _\kappa }}}{2}}}}}} \Big)\nonumber \\
 &- \frac{{{b^2} + \Delta h_d^2}}{2}F_{{\alpha _\kappa }}^\kappa \Big( {\frac{{s{C_\kappa }{G_v}}}{{{\mathcal{N}_\kappa }{{( {{b^2} + \Delta h_d^2} )}^{\frac{{{\alpha _\kappa }}}{2}}}}}} \Big),
\end{align}
\emph{when $\alpha_\kappa > 2$, $F_{{\alpha _\kappa }}^\kappa(.)$ can be expressed as}
\begin{align}
&F_{{\alpha _\kappa }}^\kappa \left( z \right){ = _2}{F_1}\big( { - \frac{2}{{{\alpha _\kappa }}},{\mathcal{N}_\kappa };1 - \frac{2}{{{\alpha _\kappa }}}; - z} \big) - 1,
\end{align}
\emph{with ${}_2{F_1}(.,.;.;.)$ being Gauss hypergeometric function.}

\emph{For most of mmWave frequencies, the path loss exponent of LOS transmissions equals to two~\cite{6824746,6655399,rappaport201238}. Therefore, when $\alpha_\kappa = 2$, $F_{{\alpha _\kappa }}^\kappa(.)$ is changed to}
\begin{align}
F_2^\kappa \left( z \right) = & \sum\limits_{c = \min \left( {1,{\mathcal{N}_\kappa } - 1} \right)}^{{\mathcal{N}_\kappa } - 1} {\frac{{\mathbf{U}\left( {{\mathcal{N}_\kappa } - 2} \right)z{N_\kappa }}}{{{{\left( {z + 1} \right)}^{{N_\kappa } - c}}\left( {{\mathcal{N}_\kappa } - c} \right)}}} \nonumber \\
&+ \frac{{{{\left( {z + 1} \right)}^{{\mathcal{N}_\kappa } - 1}} - 1}}{{{{\left( {z + 1} \right)}^{{\mathcal{N}_\kappa } - 1}}}} - z{\mathcal{N}_\kappa }\ln \Big( {1 + \frac{1}{z}} \Big),
\end{align}
\emph{$\Delta h_d=\left|h_v^{down}-h_b\right|$ and $\gamma  = \left\lfloor {r_1\sqrt {{\beta_a}{\beta_b}}} \right\rfloor$.  }
\begin{proof}
See Appendix A.
\end{proof}
\end{lemma}

\subsubsection{Coverage Probability for the Downlink Phase}
 In order to insure the quality of the downlink communicating service, a targeted rate $R^{th}_{down}$ is pre-decided in most of the cases. Then, the corresponding desired SINR threshold is given by $\Upsilon^{th}_{down}=2^{R^{th}_{down}/B_{down}}-1$, where $B_{down}$ is the bandwidth for each downlink resource block. Therefore, the coverage probability is defined as the proportion of received SINR that exceeds $\Upsilon^{th}_{down}$, which can be expressed as
\begin{align}\label{SINRPPP}
P_{down}=\mathbb{P}\left[{\Upsilon _{down}}>\Upsilon^{th}_{down}\right],
\end{align}
where $\mathbb{P}[.]$ denotes probability. With the aid of the aforementioned distance distribution and Laplace transform of interference, we present the coverage probability in the following theorem.
\begin{theorem}\label{theorem1}
\emph{When choosing the nearest UAV at $v_1$ as the corresponding UAV, then the serving communication distance is $r_1=\left\|v_1\right\|$. Accordingly, a closed-form expression for the coverage probability at the typical BS is given by}
\begin{align}\label{dmp}
&P_{down}\left( {\Upsilon _{down}^{th}} \right) \nonumber \\
\approx & \frac{\pi }{{2m\sqrt {{\beta _a}{\beta _b}} }}\sum\limits_{k = 1}^m {\sqrt {1 - {\zeta ^2}} } \sum\limits_{\gamma  \in {\mathbb{Z}^*}} {\big({p_L}\left( {\gamma \left| {h_{{v}}^{down},{h_{{b}}}} \right.} \right)}\nonumber \\
&\times{{\Psi ^L_{down}}\big( {\frac{{\zeta  + 2\gamma  + 1}}{{2\sqrt {{\beta _a}{\beta _b}} }},\Upsilon _{down}^{th}} \big)}+  { {p_N}\left( {\gamma \left| {h_{{v}}^{down},{h_{{b}}}} \right.} \right)}\nonumber \\
 & \times {\Psi _{down}^N}\big( {\frac{{\zeta  + 2\gamma  + 1}}{{2\sqrt {{\beta _a}{\beta _b}} }},\Upsilon _{down}^{th}} \big)\big),
\end{align}
\emph{where}
\begin{align}
&{\Psi ^\kappa_{down} }\left( {{r_{\rm{1}}},\Upsilon _{down}^{th}} \right) \nonumber \\
= &\sum\limits_{{n_\kappa } = 1}^{{\mathcal{N}_\kappa }} {{{\left( { - 1} \right)}^{n_\kappa + 1}}{\mathcal{N}_\kappa \choose n_\kappa}} \exp \Big( { - \frac{{{n_\kappa }{\eta _\kappa }\Upsilon _{down}^{th}n_0^2}}{{{P_v}{o_1}{C_\kappa }{{\left( {r_{\rm{1}}^2 + \Delta h_d^2} \right)}^{ - \frac{{{\alpha _\kappa }}}{2}}}}}} \Big)\nonumber \\
&\times \mathcal{L}_{down}\Big( {\frac{{{n_\kappa }{\eta _\kappa }\Upsilon _{down}^{th}}}{{{o_1}{C_\kappa }{{\left( {r_{\rm{1}}^2 + \Delta h_d^2} \right)}^{ - \frac{{{\alpha _\kappa }}}{2}}}}}\left| {{r_1}} \right.} \Big){f_n}\left( {{r_{\rm{1}}}} \right),
\end{align}
\emph{and ${\eta _\kappa } = {\mathcal{N}_\kappa }{({\mathcal{N}_\kappa }!)^{ - 1/{\mathcal{N}_\kappa }}}$. The $\zeta$ is a Gauss-Chebyshev node, which equals $\cos \left( {\frac{{2k - 1}}{{2m}}\pi } \right)|_{k=1,2,...,m}$. When $m\rightarrow \infty$, the equality holds. We are able to change the value of $m$ to balance the complexity and efficiency~\cite{7445146}. }
\begin{proof}
See Appendix B.
\end{proof}
\end{theorem}
\begin{remark}
Although the range of $\gamma$ is infinite due to $\gamma \in \mathbb{Z}^*$, we are able to choose the first three values, namely $\gamma=0,1,2$, for enhancing the computation efficiency. The reason is that when the communication distance increases, the received power of both signals and interference decreases, which have a negligible impact on the received SINR.
\end{remark}
\begin{corollary}
In the downlink phase, when considering  the inverse process where BSs send messages to UAVs, namely uplink transmissions, the coverage probability for the served UAV with a threshold $\Upsilon_{down}^{up}$ can be effortlessly deduced from \emph{\textbf{Theorem~\ref{theorem1}}}, which can be expressed as
\begin{align}
P^{up}_{down}( \Upsilon_{down}^{up} ) = P_{down}( \Upsilon_{down}^{up} )|^{\lambda_v^{down} \to \lambda_b, P_v \to P_b},
\end{align}
where $\mathcal{A} \to \mathcal{B}$ means using $\mathcal{B}$ to replace $\mathcal{A}$.
\begin{proof}
Note that the point process of UAVs is the same as that of macro BSs. Therefore, the proof procedure in \emph{\textbf{Theorem~\ref{theorem1}}} is also valid. By respectively replacing $\lambda_v^{down}$ and $P_v$ with $\lambda_b$ and $P_b$ in \emph{\textbf{Theorem~\ref{theorem1}}}, we obtain this corollary.
\end{proof}
\end{corollary}

\subsection{Performance of the Uplink Phase}

Due to the cluster property of PCP, the analysis of the uplink phase is more challenging compared with the downlink phase. Similarly, we first present the distance distribution and Laplace transform of the interference in the uplink phase. Then, the coverage performance at the corresponding UAV is analyzed.
\subsubsection{Distance Distribution in the Corresponding Cluster}
The communication distances in the same cluster with the corresponding UAV form a group $w_n=\left\|u_n^{v_0}\right\|_{n=1,2,3,...,N_{user}}$ and $u_n^{v_0} \in \mathbb{U}_{v_0}$, where $N_{user}$ is the number of users in the considered cluster. By utilizing the fair selection strategy, the subscript $n$ can be removed from $w_n$ and $u_n^{v_0}$. Regarding TCP, the PDF of the distance $(w\geq0)$ between one user in the corresponding cluster to the corresponding UAV is given by~\cite{8016632}
\begin{align}\label{PDFTho}
f_n^{Tho}( w ) = \frac{w}{{{\sigma ^2}}}\exp \big( { - \frac{{{w^2}}}{{2{\sigma ^2}}}} \big).
\end{align}

Then, for MCP, the PDF of that distance can be expressed as~\cite{7445146}
\begin{align}\label{PDFMat}
f_n^{Mat}\left( w \right) = \frac{{2w}}{{{R^2}}}\mathbf{U}\left( {R - w} \right).
\end{align}
\subsubsection{Distance Distributions in Other Clusters}
In other clusters, the distances between the users with a cluster center $v$ and the corresponding UAV compose a set $g_n=\left\|u_n^{v}\right\|_{n=1,2,3,...,N_{user}}$ and $u_n^{v} \in \mathbb{U}_{v}$. As all users are i.i.d, we are able to drop the subscript $n$ from $g_n$ and $u_n^v$. For TCP, the PDF of this distance $(g\geq0)$ is conditional on the distance $q=\left\|v\right\|\geq 0$, which is given by~\cite{8016632}
\begin{align}\label{Thoe}
f_Q^{Tho}\left( {g\left| q \right.} \right) = \frac{g}{{{\sigma ^2}}}\exp \left( { - \frac{{{g^2} + {q^2}}}{{2{\sigma ^2}}}} \right){I_0}\left( {\frac{{gq}}{{{\sigma ^2}}}} \right),
\end{align}
where $I_0(.)$ is the first kind for the modified Bessel function with order zero. Then, for MCP, the PDF of this distance is~\cite{7997055}
\begin{align}\label{Mate}
f_Q^{Mat}\left( {g\left| q \right.} \right) =& \frac{{2g}}{{\pi {R^2}}}\arccos \frac{{{g^2} + {q^2} - {R^2}}}{{2gq}}\mathbf{U}\left( {g - \left| {R - q} \right|} \right)\nonumber \\
&\times\mathbf{U}\left( {R + q - g} \right) + \frac{{2g}}{{{R^2}}}\mathbf{U}\left( {R - q - g} \right).
\end{align}
\begin{remark}
In fact, eqs. \eqref{Thoe} and \eqref{Mate} are the PDF of the distance between a daughter point to the origin. The origin can be anywhere on the plane. For example, \eqref{Thoe} and \eqref{Mate} are also valid for D2D communications, where one daughter point is fixed at the origin instead of one parent point as mentioned in the proposed system.
\end{remark}
\subsubsection{Laplace Transform of Interference in Poisson Cluster Processes}
 There exist two kinds of interference in the uplink phase. One is intra-cluster interference $I_{intra}$ and the other is inter-cluster interference $I_{inter}$. As a result, Laplace transform of interference can be defined as
\begin{align}\label{laplaceup}
\mathcal{L}_{up}(s)&=\mathbb{E}\left[\exp(-s(I_{intra}+I_{inter}))\right]\nonumber\\
&=\underbrace {\mathbb{E}\left[ {\exp ( - s{I_{intra}})} \right]}_{\mathcal{L}_a \left( s \right)}\underbrace {\mathbb{E}\left[ {\exp ( - s{I_{inter}})} \right]}_{\mathcal{L}_e \left( s \right)},
\end{align}
where
\begin{align}
&{I_{intra}} = \sum\limits_{u \in {\mathbb{U}_{{v_0}}}\backslash{u_0}} {H_g(u,h_u,h_v^{up})}, \\
&{I_{inter}} = \sum\limits_{v \in \Phi _v^{up}\backslash{v_0}} {\sum\limits_{u \in {\mathbb{U}_v}} {H_g(u,h_u,h_v^{up})} }.
\end{align}

 In several special networks, e.g., D2D communications and NOMA networks, the number of users in one cluster should be fixed as they are paired. For other networks, the number of users should be random across different clusters to enhance the generality. Therefore, we consider two cases in this paper: 1) \emph{Fixed Case}: the number of users in each cluster is fixed as $\bar N$; and 2) \emph{Random Case}: the number of users in each cluster follows Poisson distribution with the mean $\bar n$.
\begin{lemma}\label{lemma2}
\emph{When the distance between an intra-cluster interferer and the corresponding UAV is $w$, then the Laplace transform of the interference in the fixed case is given by}
\begin{align}\label{28}
&\mathcal{L}_a\left( s \right)\nonumber \\
=& {\big( {\underbrace {\sum\limits_{{\gamma _1} \in {\mathbb{Z}^*}} {\int_{\frac{{{\gamma _1}}}{{\sqrt {{\beta _a}{\beta _b}} }}}^{\frac{{{\gamma _1} + 1}}{{\sqrt {{\beta _a}{\beta _b}} }}} {\left( {{\Lambda ^L}\left( {w,s} \right) + {\Lambda ^N}\left( {w,s} \right)} \right)f_n^\Omega \left( w \right)dw} } }_{O_a\left( s \right)}} \big)^{\bar{N} - 1}},
\end{align}
\emph{where}
\begin{align}
&{\Lambda ^\kappa }\left( {w,s} \right) \nonumber \\
=& {p_\kappa }\left( {\gamma_1 \left| {{h_u},h_{{v}}^{up}} \right.} \right)\sum\limits_{i = 1}^4 {{p_i}} {\big( {1 + \frac{{s{C_\kappa }{o_i}}}{{{\mathcal{N}_\kappa }\sqrt {{{\left( {{w^2} + \Delta h_u^2} \right)}^{{\alpha _\kappa }}}} }}} \big)^{ - {\mathcal{N}_\kappa }}},
\end{align}
\emph{and $\Delta {h_u} = \left| {{h_u} - h_{{v}}^{up}} \right|$, $\gamma_1  = \left\lfloor {w\sqrt {{\beta_a}{\beta_b}}} \right\rfloor$, and $\Omega \in\{Tho,Mat\}$.}

\emph{On the other hand, if we consider the random case, the Laplace transform of the intra-cluster interference can be expressed as}
\begin{align}\label{30}
\mathcal{L}_a\left( s \right) = \exp \left( { - \left( {\bar n - 1} \right)\left( {1 - O_a\left( s \right)} \right)} \right).
\end{align}
\begin{proof}
See Appendix C.
\end{proof}
\end{lemma}
In terms of the inter-cluster interference, the Laplace transform of this interference can be expressed in the following lemma and corollary.
\begin{lemma}\label{lemma3}
\emph{When the distance between an inter-cluster interferer and the corresponding UAV is $q$, for the fixed case, the Laplace transform of inter-cluster interference can be expressed as}
\begin{align}
{\mathcal{L}_e}\left( s \right) = \exp \big( { - 2\pi \lambda _v^{up}\int_0^\infty  {\big( {1 - {{\left( {{O_e}(s,q)} \right)}^{\bar{N}}}} \big)} qdq} \big)
\end{align}
\emph{and for the random case, the corresponding Laplace transform of inter-cluster interference is changed to}
\begin{align}
&{\mathcal{L}_e}\left( s \right) \nonumber \\
=& {\exp \big( { - 2\pi \lambda _v^{up}\int_0^\infty  {\left( {1 - \exp \left( { - \bar n\left( {1 - {O_e}(s,q)} \right)} \right)} \right)} qdq} \big)},
\end{align}
\emph{where}
\begin{align}\label{33}
&{O_e}(s,q)\nonumber \\
=& \sum\limits_{{\gamma _1} \in {\mathbb{Z}^*}} {\int_{\frac{{{\gamma _1}}}{{\sqrt {{\beta _a}{\beta _b}} }}}^{\frac{{{\gamma _1} + 1}}{{\sqrt {{\beta _a}{\beta _b}} }}} {\left( {{\Lambda ^L}\left( {g,s} \right) + {\Lambda ^N}\left( {g,s} \right)} \right)f_Q^\Omega \left( {g\left| q \right.} \right)dg} }.
\end{align}
\begin{proof}
See Appendix D.
\end{proof}
\end{lemma}
\begin{remark}\label{remark4}
Note that when $A$ is large, $\exp ( -  A(1- x) )\approx x^A$, $(0\leq x \leq 1)$. Since $0\leq O_a(s) \leq 1$ and $0\leq O_e(s,q) \leq 1$, if the number of users in each cluster is large and $N=\bar n$, the difference between Laplace transform of interference in the fixed case and the random case is negligible.
\end{remark}
\subsubsection{Coverage Probability for the Uplink Phase}
 We pre-decide a targeted rate $R^{th}_{up}$ for the uplink phase such that the corresponding SINR threshold is given by $\Upsilon^{th}_{up}=2^{R^{th}_{up}/B_{up}}-1$ and $B_{up}$ is the bandwidth for each uplink resource block. Then, the coverage probability for the uplink phase is defined as
\begin{align}\label{SINRPCP}
P_{up}=\mathbb{P}\left[{\Upsilon _{up}}>\Upsilon^{th}_{up}\right].
\end{align}

Note that in the corresponding cluster, the transmitting user is randomly selected. The communication distance between the corresponding UAV and the transmitting user is $w_0=\left\|u_0\right\|$. Based on eqs. \eqref{SINRup} and \eqref{SINRPCP}, the analytical expression for the uplink coverage probability can be expressed in the following theorem and corollary.
\begin{theorem}\label{theorem2}
\emph{When the SINR threshold for the uplink phase is $\Upsilon _{up}^{th}$, then the coverage probability at the corresponding UAV is given by}
\begin{align}\label{36}
P_{up}\left( {\Upsilon _{up}^{th}} \right) \approx &\sum\limits_{{\gamma _2} \in {\mathbb{Z}^*}} {\int_{\frac{{{\gamma _2}}}{{\sqrt {{\beta _a}{\beta _b}} }}}^{\frac{{{\gamma _2} + 1}}{{\sqrt {{\beta _a}{\beta _b}} }}} {\Big( {{p_L}\left( {{\gamma _2}\left| {{h_{{u}}},h_{{v}}^{up}} \right.} \right)\Psi _{up}^L\left( {{w_0}} \right)}}} \nonumber \\
&{{{+ {p_N}\left( {{\gamma _2}\left| {{h_{{u}}},h_{{v}}^{up}} \right.} \right)\Psi _{up}^N\left( {{w_0}} \right)} \Big)d{w_0}} },
\end{align}
\emph{where}
\begin{align}
&\Psi _{up}^\kappa \left( {{w_0}} \right) \nonumber \\
=& \sum\limits_{{n_\kappa } = 1}^{{\mathcal{N}_\kappa }} {{{\left( { - 1} \right)}^{{n_\kappa } + 1}}{\mathcal{N}_\kappa \choose n_\kappa}} \exp \left( { - \frac{{{n_\kappa }{\eta _\kappa }\Upsilon _{up}^{th}n_0^2}}{{{P_u}{o_1}{C_\kappa }{{\left( {w_0^2 + \Delta h_{u}^2} \right)}^{ - \frac{{{\alpha _\kappa }}}{2}}}}}} \right)\nonumber \\
&\times \mathcal{L}_a\left( {\frac{{{n_\kappa }{\eta _\kappa }\Upsilon _{up}^{th}}}{{{o_1}{C_\kappa }{{\left( {w_0^2 + \Delta h_{u}^2} \right)}^{ - \frac{{{\alpha _\kappa }}}{2}}}}}} \right)\nonumber \\
&\times \mathcal{L}_e\left( {\frac{{{n_\kappa }{\eta _\kappa }\Upsilon _{up}^{th}}}{{{o_1}{C_\kappa }{{\left( {w_0^2 + \Delta h_{u}^2} \right)}^{ - \frac{{{\alpha _\kappa }}}{2}}}}}} \right)f_n^\Omega \left( {{w_0}} \right),
\end{align}
\emph{and $\gamma_2  = \left\lfloor {w_0\sqrt {{\beta_a}{\beta_b}}} \right\rfloor$.}
\begin{proof}
In terms of coverage probability, there exist two differences between the downlink phase and uplink phase. One is the distribution of communication distance and the other is the received interference power. Based on \emph{\textbf{Theorem~\ref{theorem1}}}, we replace \eqref{PDFPPP} by \eqref{PDFTho} and \eqref{PDFMat}. Additionally, we utilize $\mathcal{L}_a(.)\mathcal{L}_e(.)$ instead of $\mathcal{L}_{down}(.)$. After that, we obtain \emph{\textbf{Theorem~\ref{theorem2}}}.
\end{proof}
\end{theorem}
\begin{corollary}
In the uplink phase, when considering the inverse process where UAVs send messages to users, namely downlink transmissions, the coverage probability for the served user with a threshold $\Upsilon_{up}^{down}$ can be derived from \textbf{Theorem~\ref{theorem1}}, which is given by
\begin{align}
&P^{down}_{up}( \Upsilon_{up}^{down} )\nonumber \\
 =& P_{down}( \Upsilon_{up}^{down} )|^{\lambda_v^{down} \to \lambda_v^{up},f_n(.) \to f_n^\Omega(.),h_v^{down} \to h_v^{up}, h_b \to h_v}_{r_1 = 0}.
\end{align}
\begin{proof}
Note that the distribution of UAVs in the uplink phase is also a PPP and the range for interferers is $\mathbb{O}(0,+\infty)$. In contrast to \emph{\textbf{Theorem~\ref{theorem1}}}, the distance distribution of the desired communication should obey $f_n^{\Omega}(.)$ rather than $f_n(.)$. Therefore, the coverage probability for the downlink transmissions can be achieved by applying $r_1 = 0$ and $f_n(.) \to f_n^{\Omega}(.)$. Additionally, the notations of heights and densities should be modified correspondingly.
\end{proof}
\end{corollary}

By analyzing \textbf{Theorem~\ref{theorem1}} and \textbf{Theorem~\ref{theorem2}}, we are able to abstract a general expression of the coverage probability with an SINR threshold $\Upsilon _{th}$ for all point processes, including PPP and PCP.
\begin{proposition}
\emph{Assuming that Nakagami-$m$ fading channel is utilized in the considered wireless networks, then the coverage probability for both PPP and PCP can be expressed as}
\begin{align}\label{generalPc}
&{P_c}\left( {{\Upsilon _{th}}} \right) \approx \sum\limits_{{n_g} = 1}^{{m}} {{{\left( { - 1} \right)}^{{n_g} + 1}}{m \choose n_g}} \nonumber\\
&\times\int_0^\infty  {{P_{noise}}\left( {{n_g}{\Upsilon _{th}},{r_g}} \right)\mathcal{L}_I\left( {{n_g}{\Upsilon _{th}},{r_g}} \right)} {\rm{ }}f_{R_g}\left( {{r_g}} \right)d{r_g},
\end{align}
\emph{where the index $g$ represents the general case, $P_{noise}(.)$ is the signal-to-noise-ratio (SNR) coverage probability part, $\mathcal{L}_I(.)$ is the Laplace transform of interference, and $f_{R_g}(.)$ is the PDF of communication distance $r_g$ between the serving transmitter and the served receiver.}
\end{proposition}
\begin{remark}
When the Nakagami parameter $m=1$, namely Rayleigh fading channel, then the equality holds in \eqref{generalPc}.
\end{remark}

Eqs. \eqref{28}, \eqref{30}, \eqref{33}, \eqref{36} can be approximated via the Gaussian-Chebyshev quadrature equation as discussed in \eqref{dmp}. Due to the limited space, we omit it here.

\subsection{Multiple Access Scenarios}

The previous parts discussed a scenario that all transmitters are severed by a same resource block, which aims to characterize the general property of the proposed networks. In practice, multiple access techniques are used in wireless systems for supporting massive devices simultaneously. In this part, we discuss coverage probabilities under \emph{multiple access scenarios} which defined as follows: In the main beam region of receivers, all active transmitters are served in different orthogonal resource blocks, while in the side lobe region, all devices should remain silent to reduce the interference. Instead of choosing the nearest transmitter, the reference transmitter under these scenarios is randomly selected from all active transmitters in order to evaluate the average performance of networks.
\begin{corollary}
In the down phase, when considering multiple access scenarios, the coverage probability for a random selected BS in the main beam is given by
\begin{align}
P_{down}^{\rm MA}\left( {\Upsilon _{down}^{th}} \right) = {P_{up}}\left( {\Upsilon _{down}^{th}} \right)|_{\Omega  = Mat,\bar N = 1}^{\lambda _v^{up} \to {\lambda _b},h_v^{up} \to h_v^{down}}.
\end{align}
\begin{proof}
For multiple access scenarios, each BS has only one active UAV in the same resource block. The spatial framework for the uplink phase can be simplified as a Matern cluster process under a fixed case and $\bar N = 1$. The $R$ in the Matern cluster process represents the radius of BS's coverage area. With the similar proof procedure of \emph{\textbf{Theorem~\ref{theorem2}}}, we have this corollary.
\end{proof}
\end{corollary}

\noindent \textbf{Special Case 1:} Instead of serving multiple UAVs, this special case considers the scenario that each BS serves only one UAV at once. Therefore, we assume the served UAV hovers above its corresponding BS to establish transmissions.
\begin{corollary}\label{corollary1}
Under special case 1, the coverage probability can be expressed as
\begin{align}
&P_{down}^{s1}( {\Upsilon _{down}^{th}} ) \approx \sum\limits_{{n_L} = 1}^{{\mathcal{N}_L}} {{{( { - 1} )}^{{n_L} + 1}}{\mathcal{N}_L \choose n_L}} \nonumber \\
&\times \exp \Big( { - \frac{{{n_L}{\eta _L}\Upsilon _{down}^{th}n_0^2}}{{{P_v}{o_1}{C_L}\Delta h_d^{ - {\alpha _\kappa }}}}} \Big){{\cal L}_{down}}\Big( {\frac{{{n_L}{\eta _L}\Upsilon _{down}^{th}}}{{{o_1}{C_L}\Delta h_d^{ - {\alpha _\kappa }}}}| 0 } \Big).
\end{align}
\begin{proof}
By substituting $\lambda_v^{down} = \lambda_b$ and $r_1\equiv 0$ into \emph{\textbf{Theorem~\ref{theorem1}}}, we have this corollary.
\end{proof}
\end{corollary}
\begin{corollary}
In the uplink phase, when considering multiple access scenarios, the coverage probability for a random selected UAV in the main beam is given by
\begin{align}
P_{up}^{\rm MA}( {\Upsilon _{up}^{th}} ) = {P_{up}}( {\Upsilon _{up}^{th}} )|_{\bar N = 1}.
\end{align}
\begin{proof}
For multiple access scenarios, each UAV has only one active user in the same resource block. By substituting $\bar N = 1$ into \emph{\textbf{Theorem~2}}, we have this corollary.
\end{proof}
\end{corollary}

\noindent \textbf{Speical Case 2:} When a user has the highest priority among all intra-cluster users, the central UAV should fly to the location right above such user and other users should be silent or use orthogonal sub-channels.
\begin{corollary}
Under special case 2, the coverage probability in the uplink phase is changed to
\begin{align}
&P_{up}^{s2}( {\Upsilon _{up}^{th}}) \approx \sum\limits_{{n_L} = 1}^{{\mathcal{N}_L}} {{{( { - 1} )}^{{n_L} + 1}}{\mathcal{N}_L\choose n_L}} \exp \Big( { - \frac{{{n_\kappa }{\eta _\kappa }\Upsilon _{up}^{th}n_0^2}}{{{P_u}{o_1}{C_\kappa }\Delta h_{{u}}^{ - {\alpha _\kappa }}}}} \Big)\nonumber \\
&\times\exp \left( { - 2\pi \lambda _v^{up}\int_0^\infty  {\left( {1 - {O_e}\left( {\frac{{{n_\kappa }{\eta _\kappa }\Upsilon _{up}^{th}}}{{{o_1}{C_\kappa }\Delta h_{{u}}^{ - {\alpha _\kappa }}}},q} \right)} \right)} qdq} \right),
\end{align}
\begin{proof}
In special case 2, each interfering cluster has only one active user. Therefore, we use the fixed case with $\bar N=1$. Since the desired communication distance is a constant $\Delta h_{u}$, the proof procedure of this corollary is similar to \emph{\textbf{Corollary~\ref{corollary1}}}.
\end{proof}
\end{corollary}

\subsection{System Coverage Probability}
Without loss of generality, we assume that the probability of the message being successfully transmitted from the uplink to the downlink phase is $P_{s}$. It is worth noting that if the corresponding UAV is able to fly from the uplink region to the downlink region, the flight process has no impact on coverage probabilities. In the proposed system, we assume that this flight mission can be successfully completed so that $P_{s} \equiv 1$. Based on this assumption, we are capable to deduce the system coverage probability for the proposed UAV networks.
\begin{proposition}\label{proposition1}
\emph{Assuming that the SINR thresholds for the downlink phase and uplink phase are $\Upsilon _{down}^{th}$ and $\Upsilon _{up}^{th}$, respectively, then the system coverage probability is given by}
\begin{align}
P_{sc} \left( {\Upsilon _{down}^{th},\Upsilon _{up}^{th}} \right) = P_{down} \left( {\Upsilon _{down}^{th}} \right) P_{up} \left( {\Upsilon _{up}^{th}} \right)P_{s}.
\end{align}
\end{proposition}

\subsection{Practical Applications}
\begin{table*}[ht]
\newcommand{\tabincell}[2]{\begin{tabular}{@{}#1@{}}#2\end{tabular}}
\footnotesize
\centering
\caption{The Modified System Model for Three Typical UAV Applications}
\label{tab2}
\begin{tabular}{c|c|c|c}
\hline
\hline
  \textbf{Applications} &\textbf{Uplink Phase} &\textbf{Downlink Phase}  &\textbf{Cooperative Transmission}   \\ \hline
 Ubiquitous coverage &${\Phi^{up}_v}={\Phi^{down}_v}$ &\tabincell{c}{${\Phi^{down}_v}={\Phi^{up}_v}$ and \\ $b_0$ is fixed at the origin} & None, $P_{s}\equiv 1$\\  \hline
 \tabincell{c}{Information dissemination \\and data collection}  &Keep same &Keep same &Flight process, $P_{s}\equiv 1$ \\ \hline
 Relaying &Keep same  &Keep same &\tabincell{c}{$P_{s}$ is decided by the channel condition\\ of the cooperative transmission}  \\ \hline
 \hline
\end{tabular}
\end{table*}

 \begin{figure*}[th!]
  \centering
  \includegraphics[height= 2 in]{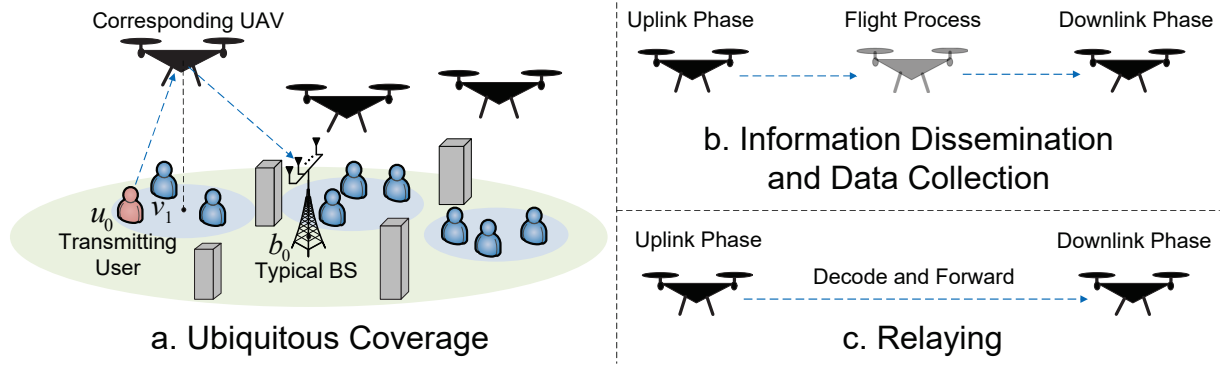}\\
  \caption{The key properties for three practical UAV-aided networks. }\label{system2}
 \end{figure*}

Next, we provide three typical UAV applications: UAV-aided ubiquitous coverage, UAV-aided information dissemination and data collection, and UAV-aided relaying~\cite{7470933}. Since all of these applications have the uplink and downlink phases in most wireless communication scenarios, the proposed unified framework can be used to model them after minor modifications. Note that we focus on a general case with limited constraints in this paper. It is able to extend to the special cases with different power allocations, user associations, beamforming patterns, and so forth, the corresponding modification processes are similar to~\cite{7445146,8016632,7862785} and hence we omit them here. As it can be seen from Fig.~\ref{system2}, for the ubiquitous coverage case, since the corresponding UAV aims to offload the data traffic of the central BS, all transceivers should be modeled in one plane. For the information dissemination and data collection case, the proposed framework can be utilized directly. For the relaying case, since two corresponding UAVs act as two relays, the success probability $P_{s}$ is mainly decided by their communication conditions.

In order to make the analysis more complete, we introduce a simplified cooperative transmission between two relays in the following part, which can be extended to other complicated scenarios. Moreover, the time slot for this transmission is located between the uplink and downlink phases. Special modifications for modeling these three applications are summarized in Table~\ref{tab2}.

\subsubsection{Simplified Cooperative Transmission}
In most of the cases, two relaying UAVs are connected through the air. Due to limited obstacles in the air, these links can be regarded as LOS transmissions. Assuming that the distance between the two relays is fixed, $y_0=\left\|v_1-v_0\right\|$ and for simplicity, the cochannel interferers can be ignored, then the received SNR at the receiver is given by
\begin{align}
\Upsilon _{link}  = {\mathrm{L} }\left( {y_0\left| {h_{{v}}^{up},h_{{v}}^{down}} \right.} \right){G_{{v_0}}}{\left| {{{\hat h}_{{v_0}}}} \right|^2}{P_v}{/}n_0^2.
\end{align}

In this cooperative transmission, $P_{s}$ represents the corresponding coverage probability, which can be calculated in the following lemma.
\begin{lemma}\label{lemma4}
\emph{When the considered cooperative transmission distance is fixed as $y_0$, the closed-form coverage probability with an SNR threshold $\Upsilon _{link}^{th}$ can be expressed as follows:}
\begin{align}
&P_{s}\left( {\Upsilon _{link}^{th}},y_0 \right)\nonumber \\
=& \frac{1}{{\left( {{\mathcal{N}_L} - 1} \right)!}}\Gamma \left( {{\mathcal{N}_L},\frac{{\mathcal{N}_L\Upsilon _{link}^{th}n_0^2{{\left( {y_0^2 + \Delta h_l^2} \right)}^{\frac{{{\alpha _L}}}{2}}}}}{{{P_v}{o_1}{C_L}}}} \right),
\end{align}
\emph{where $\Delta {h_l} = \left| {h_{{v}}^{up} - h_{{v}}^{down}} \right|$ and $\Gamma(.,.)$ is the upper incomplete gamma function.}
\begin{proof}
 By using the cumulative distribution function (CDF) of the Gamma distribution and following similar proof as in \emph{\textbf{Theorem~\ref{theorem1}}}, we obtain \emph{\textbf{Lemma~\ref{lemma4}}}.
\end{proof}
\end{lemma}
\subsubsection{Coverage Probability for the Relaying Case}
The system coverage probability for the relaying case can be represented by \textbf{Proposition~\ref{proposition1}} as well. The only difference is the probability $P_s$. Therefore, the system coverage probability is expressed as follows:
\begin{proposition}\label{proposition3}
\emph{Assuming that the SINR thresholds for the downlink phase, uplink phase, and cooperative transmission are $\Upsilon _{down}^{th}$, $\Upsilon _{up}^{th}$, and $\Upsilon _{link}^{th}$ respectively, the system coverage probability for the relaying case with a fixed cooperative transmission distance $y_0$ is given by}
\begin{align}
&P_{sc} \left( {\Upsilon _{down}^{th},\Upsilon _{up}^{th}} ,{\Upsilon _{link}^{th}},y_0\right) \nonumber\\
=& P_{down} \left( {\Upsilon _{down}^{th}} \right) P_{up} \left( {\Upsilon _{up}^{th}} \right)P_{s}\left( {\Upsilon _{link}^{th}},y_0 \right).
\end{align}
\end{proposition}
\begin{remark}\label{r7}
Note that the LOS transmissions outperform NLOS transmissions in mmWave communications. When the altitude of UAVs increases, the PDF $p_L(.)$ in \eqref{LOSp} enlarges, which enhances the system coverage probability $P_{sc}$. However, the high altitude of UAVs also increases the communication distance such that the path loss $\mathrm{L}(.)$ in \eqref{LOSorNLOS} raises. Therefore, $P_{sc}$ can be maximized by selecting optimal altitude of UAVs.
\end{remark}
\section{Numerical Results}
Since the used blockage model utilizes the bands from 3 to 60 GHz~\cite{LOSmodel}, the results can be extended to sub-6 GHz scenarios. The main difference between sub-6 GHz and mmWave is the antenna pattern and channel model. For sub-6 GHz scenarios, we use an isotropic antenna and Rayleigh fading to generate an ideal case. More specifically, we assume that $N_b=N_v=N_u=1$ and $M_c=m_c=0$ dB. We choose the notations for NLOS transmissions to represent those in sub-6 GHz scenarios, so ${{p_N}\left( {\gamma \left| {{h_t},{h_r}} \right.} \right)}  \equiv 1$ and $N_N=1$. In this section, sub-6 GHz scenarios are regarded as the benchmark.

\begin{table*}[htbp]
\centering
\footnotesize
\caption{Sectorized Antenna Pattern With Uniform Planar Square Array}
\label{tabUPA}
\begin{tabular}{l|c}
\hline
\hline
   \textbf{Number of antenna elements} & $N_c$ \\ \hline
   \textbf{Half-power beamwidth $\theta_c^a=\theta_c^e$} & $\sqrt 3 /\sqrt {{N_c}}$ \\ \hline
   \textbf{Main-lobe gain $M_c$} & $N_c$ \\ \hline
   \textbf{Side-lobe gain $m_c$} & $\frac{{\sqrt {{N_c}}  - \sqrt 3 {N_c}\sin \left( {3\pi /\left( {2\sqrt {{N_c}} } \right)} \right)/2\pi }}{{\sqrt {{N_c}}  - \sqrt 3 \sin \left( {3\pi /\left( {2\sqrt {{N_c}} } \right)} \right)/2\pi }}$  \\
\hline
\hline
\end{tabular}
\end{table*}

\subsection{Simulations and Discussions}

\begin{figure*}[t!]
\centering
\subfigure[]{\label{fig2a} \includegraphics[width= 3 in]{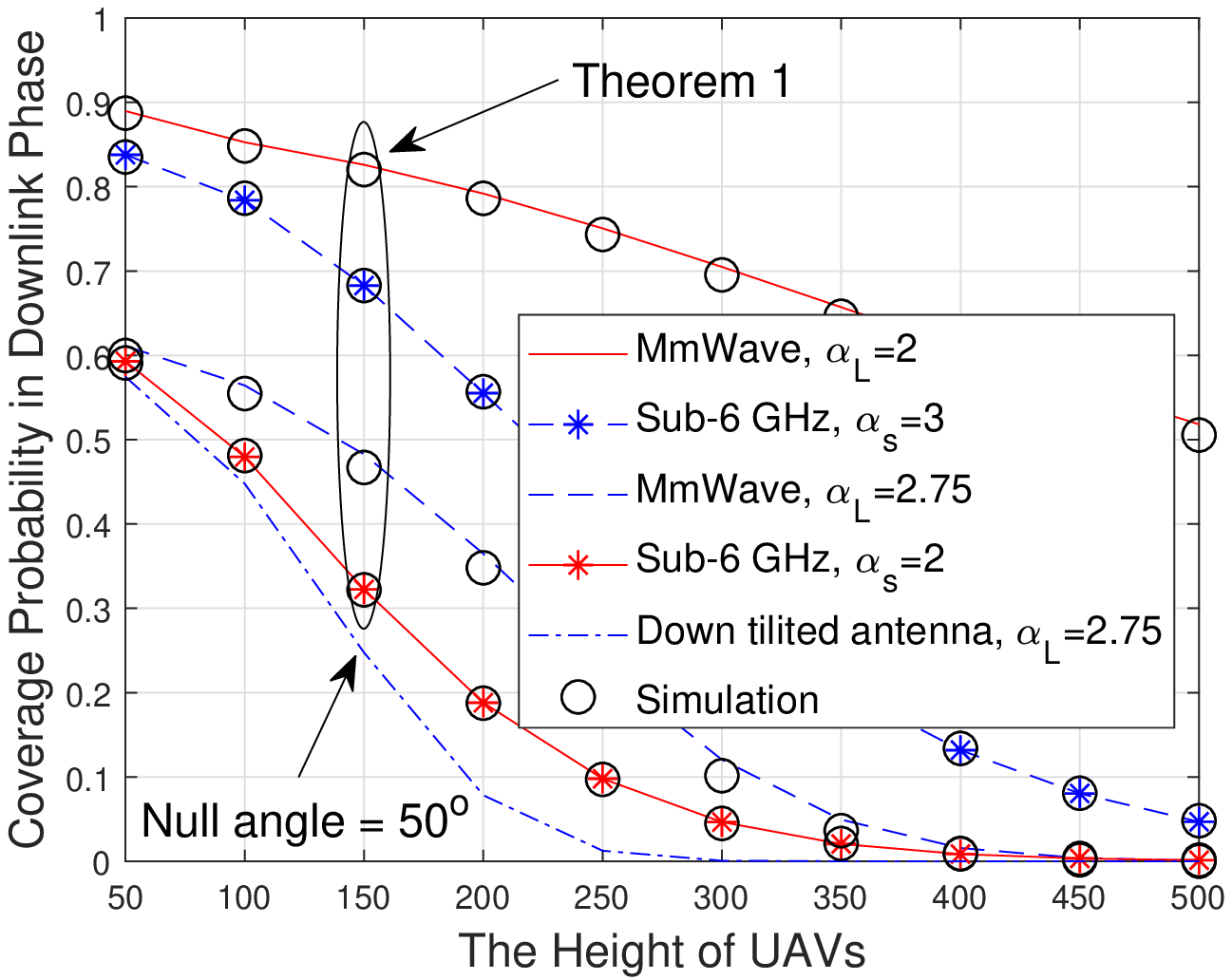}}
\subfigure[]{\label{fig2b} \includegraphics[width= 3 in]{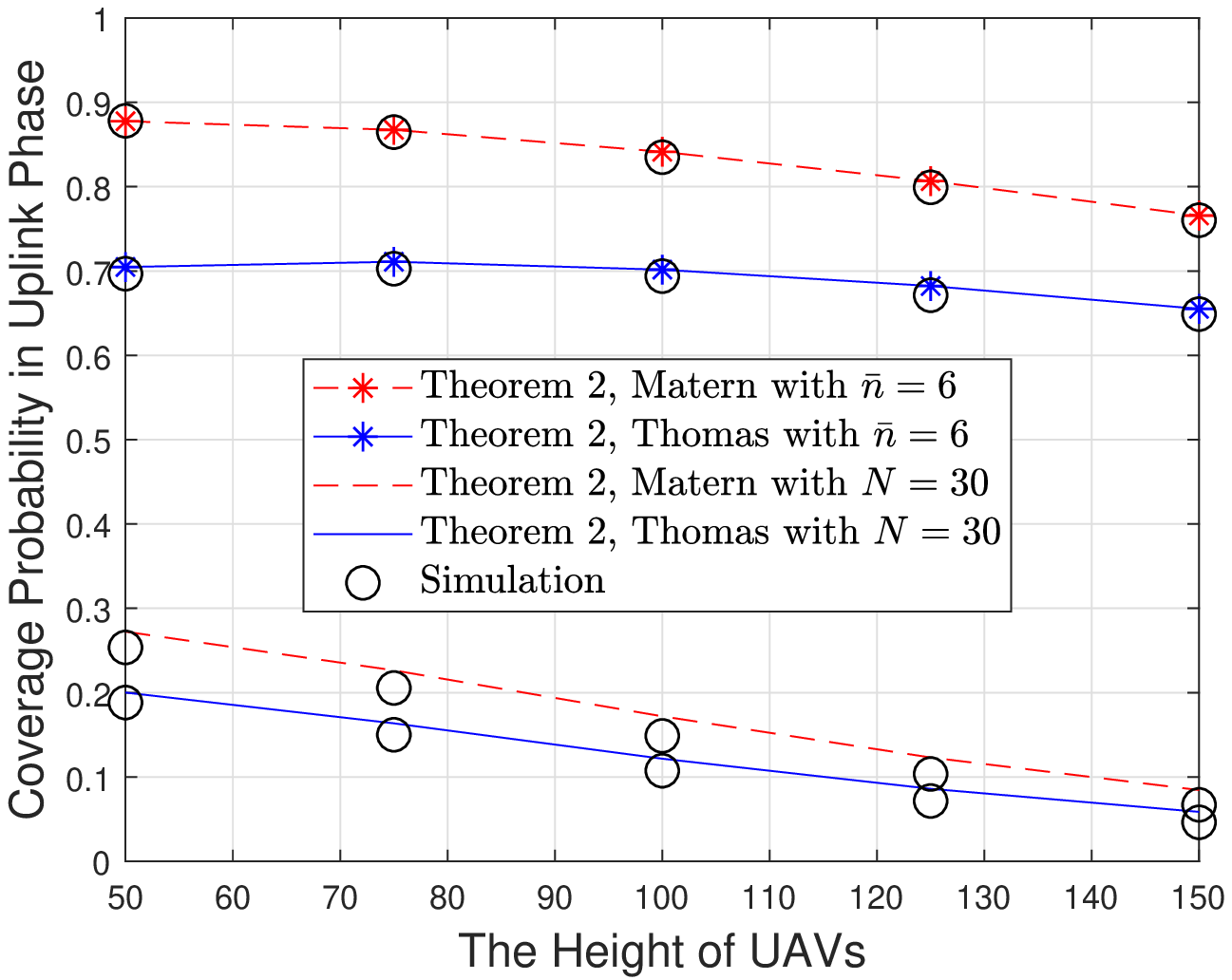}}
\caption{Monte Carlo Simulations and Validations of Coverage Probabilities with $N_b = 8$: (a) Performance in the downlink phase, with $\Upsilon^{th}_{down}=10$~dB, $h_b=10$ m for mmWave scenarios, and $\Upsilon^{th}_{down}=-10$~dB, $\lambda_v^{down}=5/ (250^2\pi)$ m$^{-2}$ for sub-6 GHz scenarios; (b) Performance in the uplink phase, with $\Upsilon^{th}_{up}=-13$~dB, the number of antenna elements $N_u=1$, and $N_v=2$.}
\label{fig2}
\end{figure*}

We use Monte Carlo (MC) simulations, which include noise, LOS and NLOS transmissions, to appraise the accuracy of derived expressions. The antenna pattern is modeled by a uniform planar square array (UPA) (see Table~\ref{tabUPA})~\cite{7491363}. General network settings are illustrated in Table~\ref{tab3}~\cite{6932503,8016632}. Additionally, we assume that $P_u=P_v=1$~W and $P_b=10$~W for simplicity. In terms of the intercept $C_\kappa$, we consider the reference distance with one meter. The validation of the derived expressions are illustrated in Fig.~\ref{fig2}, with $N_L=2$, $N_N=1$, $\beta_a=0.2$, and $\beta_b=10\times 10^{-6}$. As shown in Fig.~\ref{fig2a}, the theoretical results for the downlink phase from \textbf{Theorem~\ref{theorem1}} match MC simulations perfectly. Since antennas at BSs in traditional cellular networks are down tilted, there exists a null region right above BSs and UAVs that fly higher than BSs are frequently served by side lobes. We compare our up-tilted case with a typical down-tilted case where the null region has an elevation angle of $50^o$ and the transmitting user is the nearest one outside the null region. Fig.~\ref{fig2a} shows that although the up-tilted antenna causes further adjustments at traditional networks, it is able to provide better performance than the down-tilted scenario. Regarding the uplink phase in Fig.~\ref{fig2b}, expressions in \textbf{Theorem~\ref{theorem2}} are the tight approximations of the exact MC simulations. Particularly, the deviation is negligible in high coverage probability regions.

\begin{table*}[htbp]
\scriptsize
\centering
\caption{Network Settings of The Proposed UAV System}
\label{tab3}
\begin{tabular}{l|l|l|l}
\hline
\hline
   \textbf{Carrier frequency} & $f_{mmW}=28$ GHz, $f_{sub}=5$ GHz &\textbf{Path loss law for sub-6 GHz} & $\alpha_N=3$, $\mathcal{N}_N=1$ \\ \hline
   \textbf{Path loss law for LOS} & $\alpha_L=2$, $\mathcal{N}_L=3$   &\textbf{Path loss law for NLOS} & $\alpha_N=4$, $\mathcal{N}_N=2$\\ \hline
   \textbf{Density} & $\lambda_b=\lambda_v^{down}=\lambda_v^{up}=1/ (250^2\pi)$ m$^{-2}$ &\textbf{Altitude} &$h_u=0$ m, $h_b=30$ m\\ \hline
   \textbf{Bandwidth for mmWave} & $B_{down}=B_{up}=100$ MHz &\textbf{Radius for MCP} & $R=100$\\ \hline
   \textbf{Bandwidth for sub-6 GHz} & $B_{down}=B_{up}=10$ MHz   &\textbf{Standard deviation for TCP} & $\sigma=100$\\ \hline
   \textbf{Building density} & $\beta_a=0.5$, $\beta_b=300\times 10^{-6}$   & \textbf{Scale parameter for buildings}   & $\varepsilon=20$ m \\
\hline
\hline
\end{tabular}
\end{table*}

\subsection{Comparison and Analysis of MCP and TCP}

\begin{table*}[htbp]
\centering
\footnotesize
\caption{Path Loss Exponents for Different Carrier Frequencies in MmWave Scenarios}
\label{tab4}
\begin{tabular}{c|c|c|c}
\hline
\hline
   \textbf{Carrier frequencies}    & 28G    &38G    &60G    \\ \hline
   \textbf{LOS $\alpha_L$ }    & 2    &2   &2.25      \\ \hline
   \textbf{Strongest NLOS $\alpha_N$}     & 3    &3.71    &3.76       \\ \hline
   \textbf{Number of antenna elements}  &$N_c=4\times 4$ &$N_c=8\times 8$ &$N_c=12 \times 12$ \\ \hline
   \hline
\end{tabular}
\end{table*}

\begin{figure*}[t!]
\centering
\subfigure[]{\label{fig3a} \includegraphics[width= 3 in]{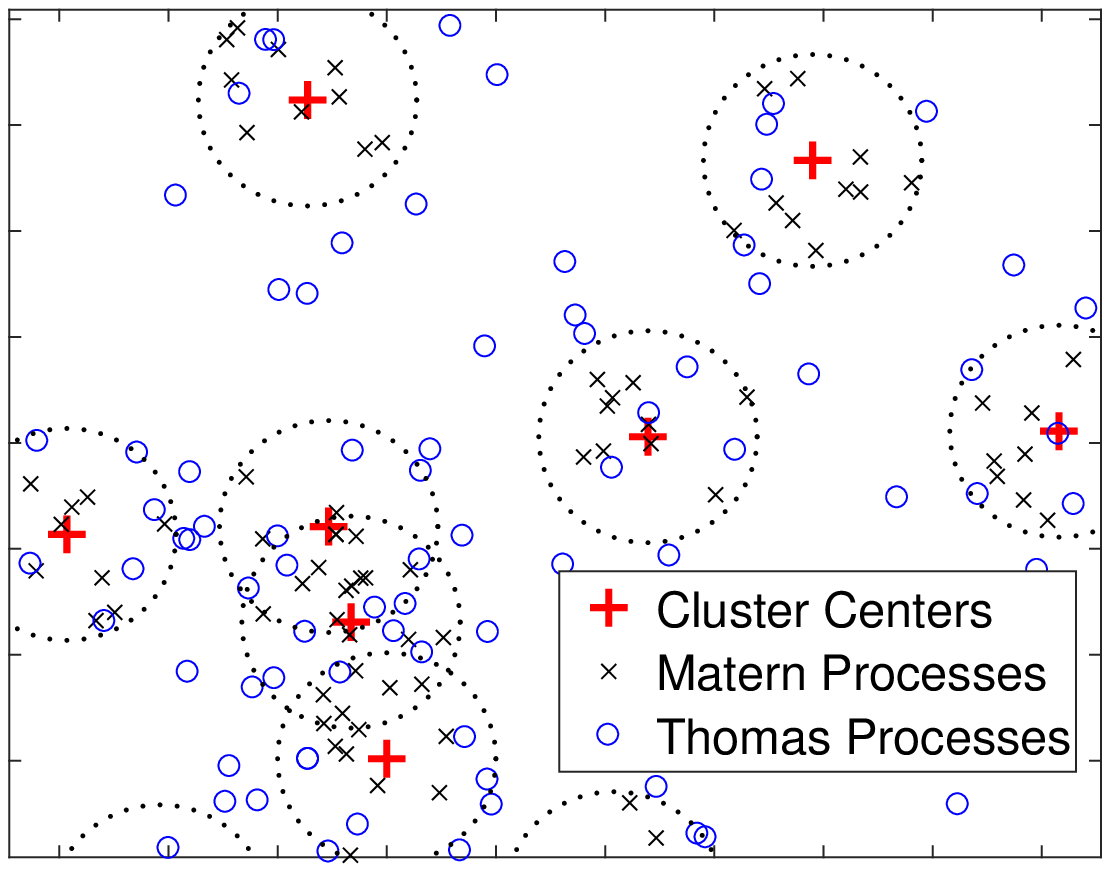}}
\subfigure[]{\label{fig3b} \includegraphics[width= 3 in]{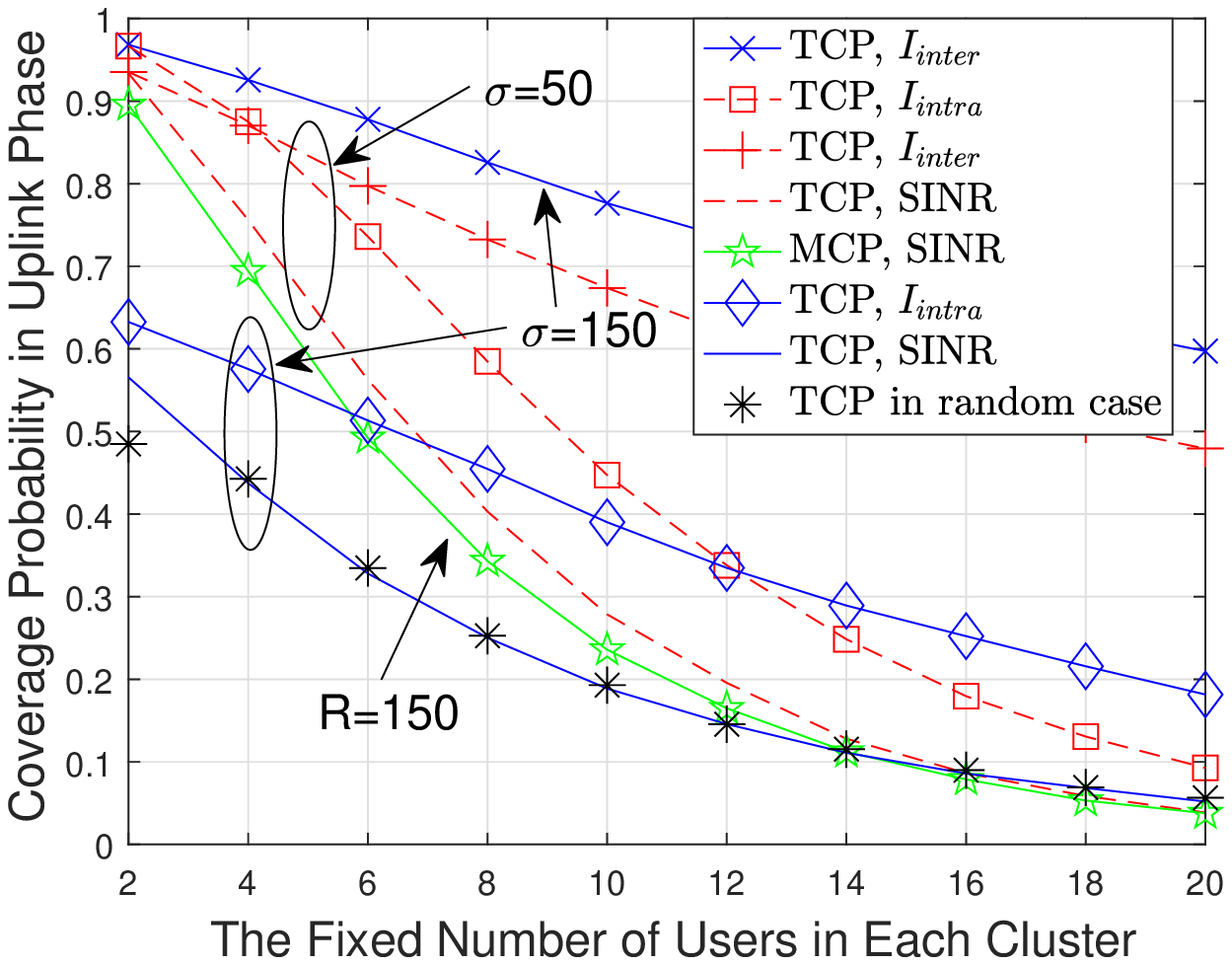}}
\caption{Comparison and Analysis of MCP and TCP: (a) Comparing the spatial difference between MCP and TCP, with the fixed number of daughter points in each cluster $\bar N=10$; (b) Coverage probability in the uplink phase versus the fixed number of users in each cluster $\bar N$, with $N_b = 8$, $N_v = 4$, $N_u = 2$, and $\Upsilon^{th}_{up}=0$~dB.}
\label{fig5}
\end{figure*}

 Regarding the clustered property of the proposed framework, we compare two typical PCPs (MCP and TCP) in Fig.~\ref{fig5}. More specifically,  Fig.~\ref{fig3a} demonstrates that the clusters in MCP are appropriate to model bounded regions, while the counterparts in TCP are suitable for open-boundary scenarios. In addition, Fig.~\ref{fig3b} shows that the coverage probability in the uplink phase has a negative correlation with the number of users in one cluster. In terms of intra-cluster and inter-cluster interference, exact results can be approximated by the simplified scenario that only considers intra-cluster interference, which significantly improve the analysis efficiency. When $\sigma$ increases from 50 to 150, the effect of inter-cluster interference is gradually enhanced. Moreover, when $R=\sigma$, coverage probabilities with MCP are higher than those with TCP in small $\bar N$ regions. Although the number of users in each cluster can be a constant $\bar N$ or a variable with the mean $\bar n$, coverage probabilities for two cases are nearly the same under the condition $\bar N=\bar n$ in high $\bar N$ regions, which indicates that two cases can be replaced by each other as discussed in \textbf{Remark~\ref{remark4}}.

\subsection{The Impact of Blockage Environment and Noise}

\begin{figure*}[t!]
\centering
\subfigure[]{\label{fig4a} \includegraphics[width= 3 in]{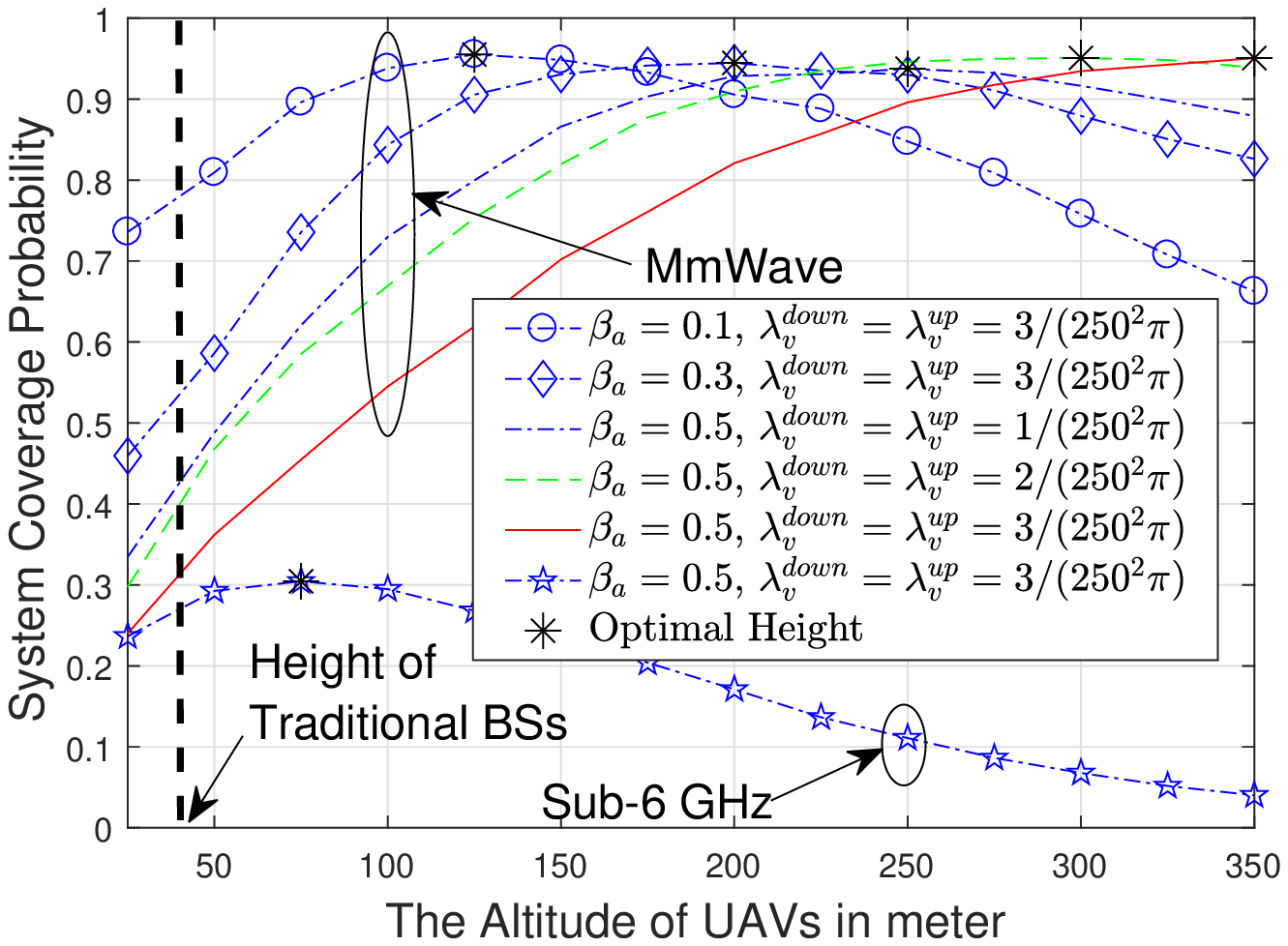}}
\subfigure[]{\label{fig4b} \includegraphics[width= 3 in]{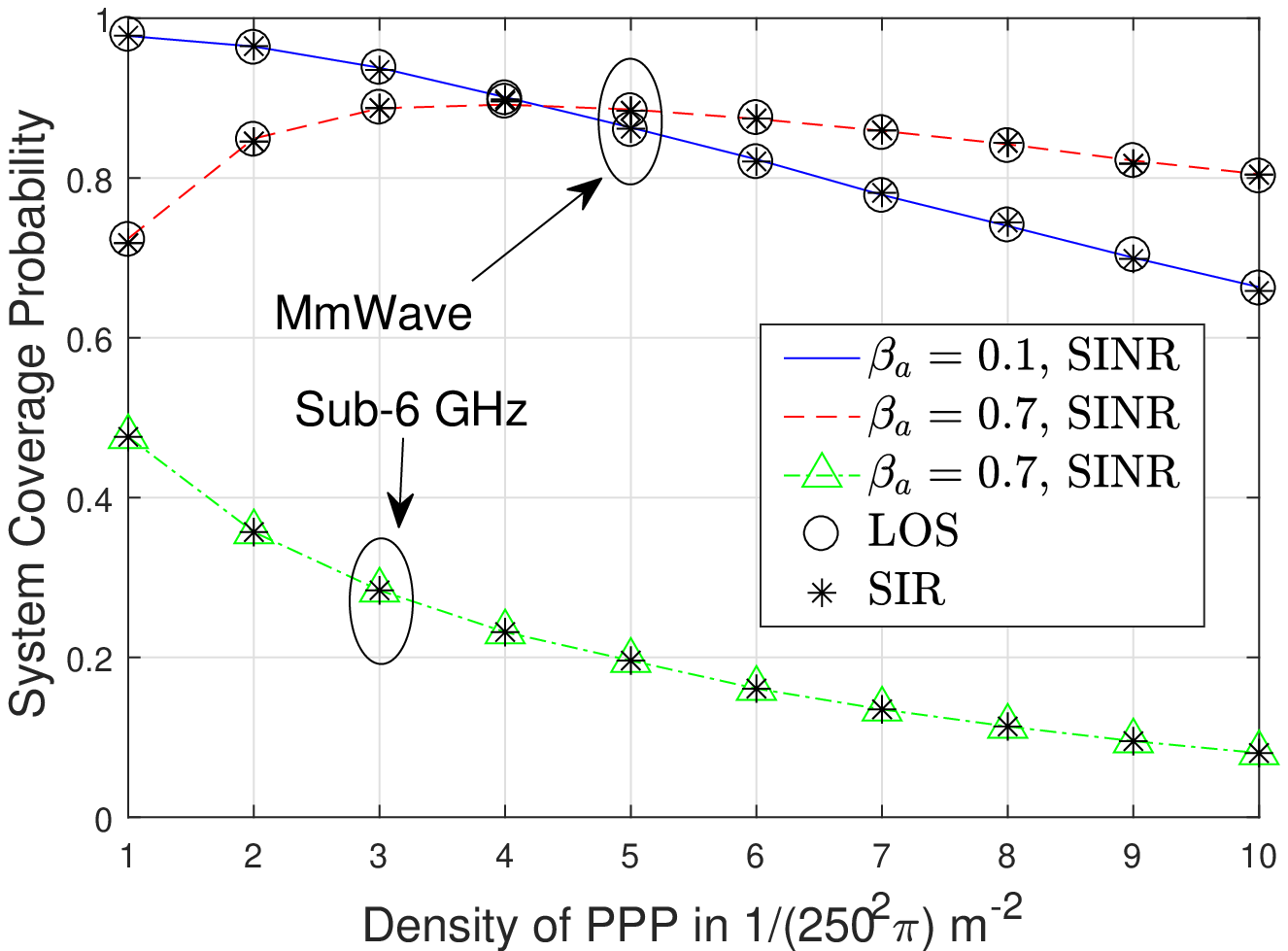}}
\caption{Impact of Blockage Environment and Thermal Noise with $N_b = 8$, $N_v = 4$, and $N_u = 2$: (a) System coverage probability versus the altitude of UAVs, with $y_0=250$ m; (b) System coverage probability versus the density of PPP ($\lambda_v^{up}=\lambda_v^{down}$),  with $y_0=200$ m and $\beta_b=100 \times 10^{-6}$.}
\label{fig6}
\end{figure*}

 In this part, we analyze the system coverage performance affected by the blockage environment and the thermal noise, with $\bar N=30$, $h_v=100$ m, $\Upsilon^{th}_{up}=\Upsilon^{th}_{down}=\Upsilon^{th}_{link}=-20$~dB, and the bandwidth for all scenarios equals $100$ MHz. Since MCP has similar trends with TCP and thus we only study TCP here. Regarding blockage effects, Fig.~\ref{fig4a} shows that an optimal altitude of UAVs exists for maximizing the system coverage probability as discussed in \textbf{Remark~\ref{r7}}. It is interesting that all optimal values are larger than the height of traditional BSs, which is around 30 m (dash line in Fig.~\ref{fig4a}). Therefore, UAV networks are capable of achieving better performance than terrestrial cellular networks by adjusting the serving altitude. When the density of UAVs increases from $\lambda_v^{up}=\lambda_v^{down}=1/(250^2\pi)$ to $3/(250^2\pi)$ m$^{-2}$, the optimal altitude rises. Moreover, the hight density of obstacles $\beta_a$ also enlarges the optimal height. In terms of signal-to-interference-ratio (SIR), Fig.~\ref{fig4b} demonstrates that the thermal noise has a negligible effect on system coverage probabilities in both mmWave and sub-6 GHz scenarios. NLOS transmissions can be ignored in mmWave communications in the considered framework. Additionally, both figures in Fig.~\ref{fig6} illustrate that mmWave outperforms sub-6 GHz.

\subsection{The Impact of Antennas and Carrier Frequencies in MmWave Scenarios}
\begin{figure*}[t!]
\centering
\subfigure[]{\label{fig5a} \includegraphics[width= 3 in]{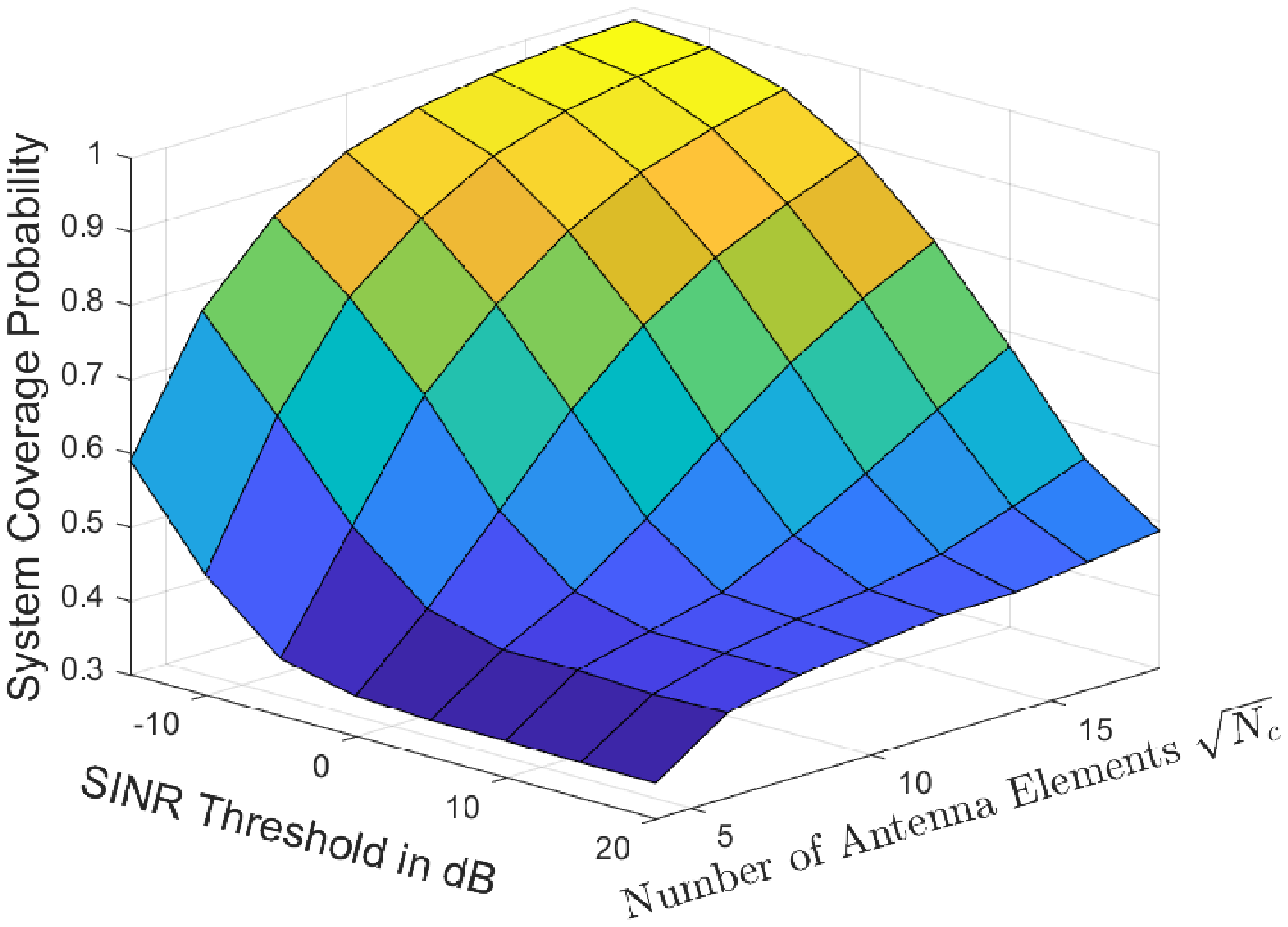}}
\subfigure[]{\label{fig5b} \includegraphics[width= 3 in]{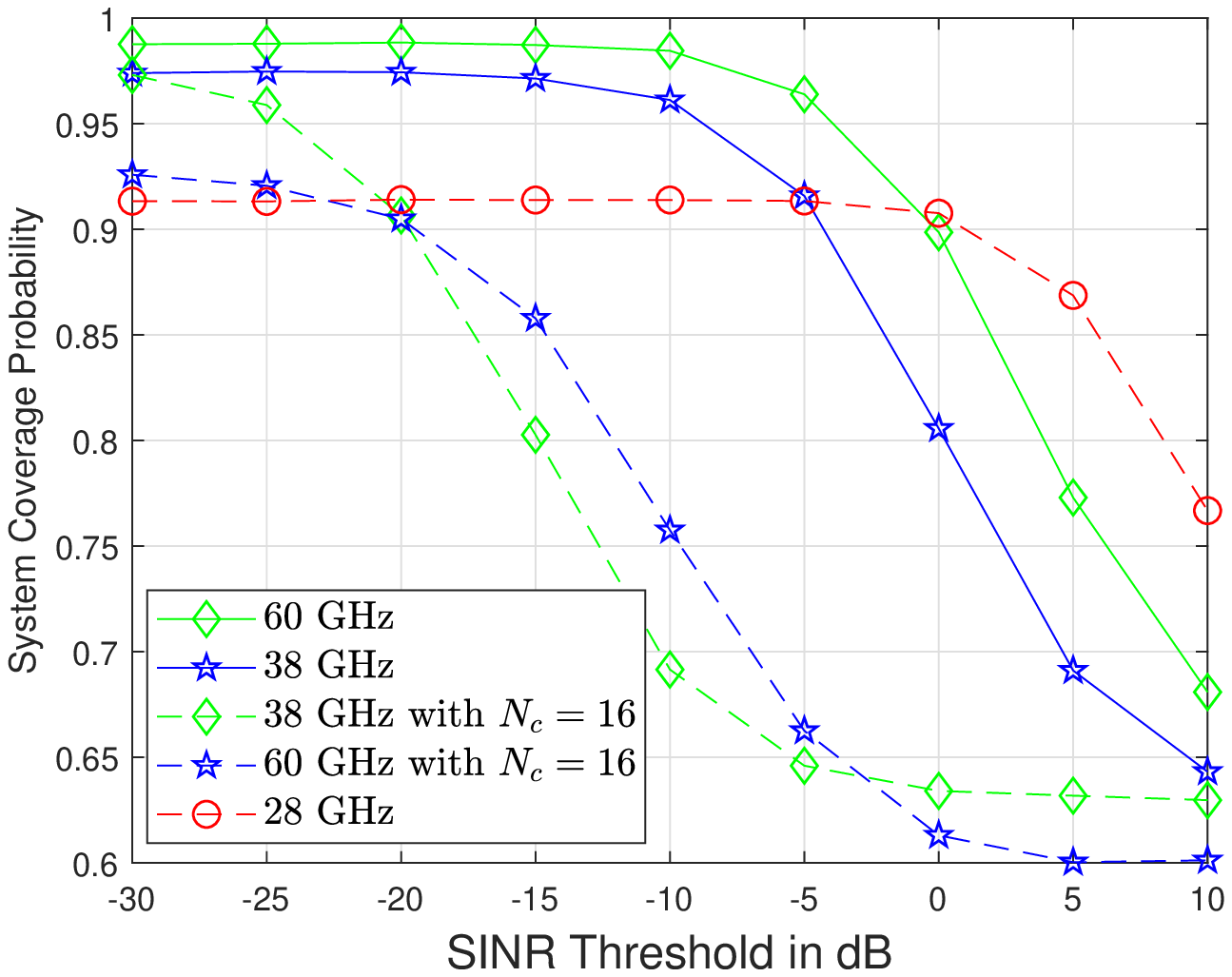}}
\caption{Impact of Antenna Scale and Different MmWave Frequencies: (a) System coverage probability versus the number of antenna elements and SINR threshold, with TCP, $\bar N=8$, $y_0=200$ m, and $h_v=100$ m; (b) System coverage probability versus SINR threshold, with TPC, $\beta_a = 0.3$, $\bar N=2$, $y_0=250$ m, and $h_v=100$ m.}
\end{figure*}

Since sub-6 GHz can be regarded as one special case of mmWave scenarios, we study mmWave scenarios in this part to comprehensively evaluate the proposed framework. Fig.~\ref{fig5a} concentrates on the impact of antenna scales. For UPA, the large number of antenna elements is capable of increasing the main lobe gain and narrowing the main lobe beamwidth. Therefore, the desired signal is enhanced and the interference is weakened. The system coverage probability has a positive correlation with the number of antenna elements $N_c$ as shown in Fig.~\ref{fig5a}. Then, we focus on different carrier frequencies. It is worth noting that compared with 28 GHz, higher carrier frequencies allows more antenna elements to be deployed at transceivers. We provide the path loss laws and the estimated number of antennas of three typical carrier frequencies for mmWave communications in Table~\ref{tab4}~\cite{6655399,rappaport201238}. The comparison of those candidates are illustrated in Fig.~\ref{fig5b}. When the number of antennas is fixed as 16, namely $N_c=4\times 4 =16$, 28 GHz with the best path loss law and the intercept achieves the highest coverage probabilities in high SINR threshold region, while 38 GHz outperforms other two candidates in low SINR threshold area. With the increasing of the antenna scale, 60 GHz becomes the best among three candidates.

\section{Conclusion}

 In this article, a unified 3D spatial framework for UAV networks has been provided, where the stochastic geometry has been utilized for modeling the locations of BSs, UAVs and users. Especially, in the uplink phase, two typical PCPs (MCP and TCP) have been analyzed to enhance the generality. Tractable expressions in terms of coverage probabilities have been deduced for mmWave communications, which can be extended to sub-6 GHz scenarios. For the proposed system, there exists an optimal altitude of UAVs to achieve the maximum coverage probability. Another remark is that the effects of the thermal noise and NLOS transmissions are negligible in mmWave-aided UAV networks. Moreover, large number of antenna elements have the capability of enhancing the coverage performance. Without considering the antenna scales, 28 GHz is the best choice for mmWave scenarios in low SINR regions. Since practical antennas with multiple beams are more accurate than the proposed sectorized model and the optimal location of UAVs exists for different applications, we will discuss the actual antenna effect and the location optimization in our future work.

\numberwithin{equation}{section}
\section*{Appendix~A: Proof of Lemma~\ref{lemma1}} \label{Appendix:A}
\renewcommand{\theequation}{A.\arabic{equation}}
\setcounter{equation}{0}
When substituting $I$ into \eqref{laplace}, the Laplace transform of interference in the downlink phase can be written as
\begin{align}\label{A.1}
&\mathcal{L}_ {down}^{mmW}\left( {s\left| {{r_1}} \right.} \right) \nonumber \\
=& \mathbb{E}\Big[ {\exp \big( { - s\sum\limits_{v \in \Phi _v^{down}\backslash{v_1}} {{}\mathrm{L}\left( {\left\| v \right\|\left| {h_v^{down},{h_{{b_0}}}} \right.} \right){G_v}{{\left| {{{\hat h}_v}} \right|}^2}} } \big)} \Big].
\end{align}

We introduce $V_L$ to represent LOS transmissions, which obeys $V_L=\mathbb{B}\left( {{p_L}\left( {\gamma \left| {{h^{down}_v},{h_b}} \right.} \right)} \right)$. Assuming that the interfering distance $r=\left\|v\right\|$, with the aid of \eqref{LOSorNLOS} and \eqref{A.1}, Laplace transform of interference via LOS links $\Xi_L$ \eqref{A.2} can be expressed at the top of next page.
\begin{figure*}[ht]
\normalsize
\begin{align}\label{A.2}
{\Xi _L} = &\mathbb{E}\bigg[ {\exp \bigg( { - \frac{{s\sum\limits_{_{v \in \Phi _v^{down}/{v_1}}} {{C_L}{G_v}{{\left| {{{\hat h}_v}} \right|}^2}V_L } }}{{\sqrt {{{\left( {{r^2} + {{\left| {h_v^{down} - {h_b}} \right|}^2}} \right)}^{{\alpha _L}}}} }}} \bigg)} \bigg] \mathop  = \limits^{\left( a \right)} \mathbb{E}\bigg[ {\prod\limits_{_{v \in \Phi _v^{down}\backslash{v_0}}} {{\mathbb{E}_v}\bigg[ {{{\bigg( {1 + \frac{{s{C_L}{G_v}V_L }}{{{N_L}\sqrt {{{\left( {{r^2} + {{\left| {h_v^{down} - {h_b}} \right|}^2}} \right)}^{{\alpha _L}}}} }}} \bigg)}^{ - {N_L}}}} \bigg]} } \bigg]\displaybreak[4] \nonumber \\
\mathop  = \limits^{\left( b \right)}& \exp \Bigg( { - 2\pi \lambda _v^{down}{\mathbb{E}_{{G_v}}}\left[ {{\mathbb{E}_{V_L} }\left[ {\int_{{r_1}}^\infty  {\left( {1 - {{\left( {1 + \frac{{s{C_L}{G_v}V_L }}{{{N_L}\sqrt {{{\left( {{r^2} + \Delta h_d^2} \right)}^{{\alpha _L}}}} }}} \right)}^{ - {N_L}}}} \right)rdr} } \right]} \right]} \Bigg),
\end{align}
\hrulefill \vspace*{0pt}
\end{figure*}
In \eqref{A.2}, (a) computes the expectation of the normalized Gamma variable $| {{{\hat h}_v}} |^2$. (b) follows the fact that interfering UAVs are i.i.d. as a PPP and located further than the serving UAV at $v_1$.

When $\alpha_\kappa>2$, a closed-form expression is shown as follows:
\begin{align}\label{A.3}
&{\rm{Z}}_{{N_\kappa }}^\kappa \left( {s,a,b,{G_v}} \right) \nonumber \\
=& \int_a^b {\left( {1 - {{\left( {1 + \frac{{s{C_L}{G_v}}}{{{N_L}\sqrt {{{\left( {{r^2} + \Delta h_d^2} \right)}^{{\alpha _L}}}} }}} \right)}^{ - {N_L}}}} \right)rdr} \nonumber \\
\mathop  = \limits^{\left( c \right)}& \int_{{{\left( {{a^2} + \Delta h_d^2} \right)}^{\frac{1}{2}}}}^{{{\left( {{b^2} + \Delta h_d^2} \right)}^{\frac{1}{2}}}} {\left( {1 - {{\left( {1 + \frac{{s{C_L}{G_v}}}{{{N_L}{y^{{\alpha _L}}}}}} \right)}^{ - {N_L}}}} \right)ydy} \nonumber\\
\mathop  = \limits^{\left( d \right)}& \frac{{{a^2} + \Delta h_d^2}}{2}F_{{\alpha _\kappa }}^\kappa \left( {\frac{{s{C_\kappa }{G_v}}}{{{N_\kappa }{{\left( {{a^2} + \Delta h_d^2} \right)}^{\frac{{{\alpha _\kappa }}}{2}}}}}} \right)- \frac{{{b^2} + \Delta h_d^2}}{2} \nonumber \\
&\times F_{{\alpha _\kappa }}^\kappa \left( {\frac{{s{C_\kappa }{G_v}}}{{{N_\kappa }{{\left( {{b^2} + \Delta h_d^2} \right)}^{\frac{{{\alpha _\kappa }}}{2}}}}}} \right),
\end{align}
where (c) follows $y = {\left( {{r^2} + \Delta h_d^2} \right)^{\frac{1}{2}}}$. (d) depends on one integral $\int_A^\infty  {( {1 - {{( {1 + \tilde s{y^{ - \alpha }}} )}^{ - N}}} )yd} y = \frac{{{A^2}}}{2}\left( {_2{F_1}\left( { - \frac{2}{\alpha },N;1 - \frac{2}{\alpha }; - \frac{{\tilde s}}{{{A^\alpha }}}} \right) - 1} \right)$~\cite{7875124,8401954}. When $\alpha_\kappa=2$, the closed-form expression can be expressed as follows:
\begin{align}\label{A.4}
{\rm{Z}}_{{N_\kappa }}^\kappa \left( {s,a,b,{G_v}} \right)\mathop  = \limits^{\left( e \right)} &\frac{{{a^2} + \Delta h_d^2}}{2}F_{{2 }}^\kappa \left( {\frac{{s{C_\kappa }{G_v}}}{{{N_\kappa }{{\left( {{a^2} + \Delta h_d^2} \right)}^{\frac{{{\alpha _\kappa }}}{2}}}}}} \right) \nonumber\\
&- \frac{{{b^2} + \Delta h_d^2}}{2}F_{{2 }}^\kappa \left( {\frac{{s{C_\kappa }{G_v}}}{{{N_\kappa }{{\left( {{b^2} + \Delta h_d^2} \right)}^{\frac{{{\alpha _\kappa }}}{2}}}}}} \right),
\end{align}
where (e) follows (2.117-1), (2.117-3) and (2.118-1) in~\cite{jeffrey2007table} for $N_\kappa>1$ and (2.118-2) in ~\cite{jeffrey2007table} for $N_\kappa=1$. By substituting \eqref{A.3} and \eqref{A.4} into \eqref{A.2}, we obtain
\begin{align}\label{A.5}
{\Xi _L} \mathop= \limits^{\left( f \right)}& \exp \Big( - 2\pi \lambda _v^{down}{\mathbb{E}_{{G_v}}}\Big[{p_L}\left( {\gamma \left| {h_v^{down},{h_b}} \right.} \right)\nonumber\\
&\times {\rm{Z}}_{{N_L}}^L\left( {s,{r_1},\frac{{\gamma  + 1}}{{\sqrt {{\beta _a}{\beta _b}} }},{G_v}} \right)+ \sum\limits^\infty_{j = \gamma  + 1}  {{p_L}\left( {j\left| {h_v^{down},{h_b}} \right.} \right)}\nonumber \\
&{\times{\rm{Z}}_{{N_L}}^L\left( {s,\frac{j}{{\sqrt {{\beta _a}{\beta _b}} }},\frac{{j + 1}}{{\sqrt {{\beta _a}{\beta _b}} }},{G_v}} \right)} \Big]\Big)\nonumber \\
\mathop  = \limits^{\left( g \right)}& \exp \left( { - 2\pi \lambda _v^{down}\sum\limits_{i = 1}^4 {{p_i}} {\Theta ^L}\left( {s,{o_i}\left| {{r_1}} \right.} \right)} \right),
\end{align}
where (f) depends on step property of \eqref{LOSp}. (g) calculates the expectation of the variable $o_i$ for antenna gain. Then, with the similar process, Laplace transform of interference via NLOS links $\Xi_N$ is given by
\begin{align}\label{A.6}
{\Xi _N} = \exp \big( { - 2\pi \lambda _v^{down}\sum\limits_{i = 1}^4 {{p_i}} \Theta^N\left( {s,{o_i}\left| {{r_1}} \right.} \right)} \big).
\end{align}

\numberwithin{equation}{section}
\section*{Appendix~B: Proof of Theorem~\ref{theorem1}} \label{Appendix:B}
\renewcommand{\theequation}{B.\arabic{equation}}
\setcounter{equation}{0}

Based on the eq. \eqref{SINRPPP}, we first derive the coverage probability for mmWave scenarios, which can be divided into LOS parts and NLOS parts:
\begin{align}\label{B.1}
P_{down}^{mmW}\left( {\Upsilon _{down}^{th}} \right) = &\mathbb{P}\left[ {\Upsilon _{down}^{mmW} > \Upsilon _{down}^{th}} \right]\nonumber \\
=& P_{down}^L\left( {\Upsilon _{down}^{th}} \right) + P_{down}^N\left( {\Upsilon _{down}^{th}} \right).
\end{align}

For LOS links, the corresponding coverage probability \eqref{B.2} can be expressed at the top of next page.
\begin{figure*}[!t]
\normalsize
\begin{align}\label{B.2}
P_{down}^L\left( {\Upsilon _{down}^{th}} \right) = &\mathbb{P}\left[ {\frac{{V_L {C_L}{{\left( {r_{\rm{1}}^2 + \Delta h_d^2} \right)}^{ - \frac{{{\alpha _L}}}{2}}}{G_{{0}}}{{\left| {{{\hat h}_{{v_1}}}} \right|}^2}}}{{\left( {{I_{down}} + {{n_0^2} \mathord{\left/
 {\vphantom {{n_0^2} {{P_v}}}} \right.
 \kern-\nulldelimiterspace} {{P_v}}}} \right)}} > \Upsilon _{down}^{th}\left| {{r_{\rm{1}}} = \left\| {{v_{\rm{1}}}} \right\|} \right.} \right]\nonumber \\
 \mathop  = \limits^{(a)}& \int_0^\infty  {\mathbb{P}\left[ {{{\left| {{{\hat h}_{{v_1}}}} \right|}^2} > \frac{{\Upsilon _{down}^{th}\left( {{I_{down}} + {{n_0^2} \mathord{\left/
 {\vphantom {{n_0^2} {{P_v}}}} \right.
 \kern-\nulldelimiterspace} {{P_v}}}} \right)}}{{V_L {C_L}{G_{{0}}}{{\left( {r_{\rm{1}}^2 + \Delta h_d^2} \right)}^{ - \frac{{{\alpha _L}}}{2}}}}}} \right]} {f_n}\left( {{r_{\rm{1}}}} \right)d{r_{\rm{1}}}\nonumber \\
 \mathop  \approx \limits^{(b)}& \int_0^\infty  {\left( {1 - \mathbb{E}\left[ {{{\left( {1 - \exp \left( { - \frac{{{\eta _L}\Upsilon _{down}^{th}\left( {{I_{down}} + {{n_0^2} \mathord{\left/
 {\vphantom {{n_0^2} {{P_v}}}} \right.
 \kern-\nulldelimiterspace} {{P_v}}}} \right)}}{{V_L {C_L}{G_{{0}}}{{\left( {r_{\rm{1}}^2 + \Delta h_d^2} \right)}^{ - \frac{{{\alpha _L}}}{2}}}}}} \right)} \right)}^{{N_L}}}} \right]} \right){f_n}\left( {{r_{\rm{1}}}} \right)d{r_{\rm{1}}}} \nonumber \\
 \mathop  = \limits^{(c)}& \sum\limits_{\gamma  \in {\mathbb{Z}^*}} {{p_L}\left( {\gamma \left| {h_{{v_1}}^{down},{h_{{b_0}}}} \right.} \right)\sum\limits_{{n_L} = 1}^{{N_L}} {{{\left( { - 1} \right)}^{n + 1}}{N_L \choose n_L}} }{\int_{\frac{\gamma }{{\sqrt {{\beta _a}{\beta _b}} }}}^{\frac{{\gamma  + 1}}{{\sqrt {{\beta _a}{\beta _b}} }}} {\exp \left( { - \frac{{{n_L}{\eta _L}\Upsilon _{down}^{th}n_0^2}}{{{P_v}{o_1}{C_L}{{\left( {r_{\rm{1}}^2 + \Delta h_d^2} \right)}^{ - \frac{{{\alpha _L}}}{2}}}}}} \right)} } \nonumber \\
 & \times \mathbb{E}\left[ {\exp \left( { - \frac{{{n_L}{\eta _L}\Upsilon _{down}^{th}{I_{down}}}}{{{o_1}{C_L}{{\left( {r_{\rm{1}}^2 + \Delta h_d^2} \right)}^{ - \frac{{{\alpha _L}}}{2}}}}}} \right)} \right]{f_n}\left( {{r_{\rm{1}}}} \right)d{r_{\rm{1}}}\nonumber \\
 \mathop  = \limits^{(d)} &\frac{\pi }{{2m\sqrt {{\beta _a}{\beta _b}} }}\sum\limits_{k = 1}^m {\sqrt {1 - {\zeta ^2}} } \sum\limits_{\gamma  \in {\mathbb{Z}^*}} {{p_L}\left( {\gamma \left| {h_{{v_1}}^{down},{h_{{b_0}}}} \right.} \right){\Psi _{down}^L}\left( {\frac{{\zeta  + 2\gamma  + 1}}{{2\sqrt {{\beta _a}{\beta _b}} }},\Upsilon _{down}^{th}} \right)},
\end{align}
\hrulefill \vspace*{0pt}
\end{figure*}
In \eqref{B.2}, (a) follows the fact that the serving UAV is the closest node to the reference BS. (b) depends on a tight upper bound for the normalized gamma variable $|\hat {h_{v_1}}|^2$, which is $\mathbb{P}\left[ {|\hat {h_{v_1}}|^2 < \chi } \right] \le {\left( {1 - {e^{ - {\eta _L}\chi }}} \right)^{{N_L}}}$~\cite{alzer1997some}. (c) calculates the expectation of LOS variable $V_L$ and the directional antenna gain is $o_1$ for the reference communication. (d) bases on the Gaussian-Chebyshev quadrature equation~\cite{7445146}. With the similar deriving procedure, the coverage probability for NLOS transmissions is given by
\begin{align}\label{B.3}
&P_{down}^N\left( {\Upsilon _{down}^{th}} \right) \approx \frac{\pi }{{2m\sqrt {{\beta _a}{\beta _b}} }}\sum\limits_{k = 1}^m {\sqrt {1 - {\zeta ^2}} } \nonumber\\
&\times\sum\limits_{\gamma  \in {\mathbb{Z}^*}} {{p_N}\left( {\gamma \left| {h_{{v_1}}^{down},{h_{{b_0}}}} \right.} \right)} {\Psi _{down}^N}\left( {\frac{{\zeta  + 2\gamma  + 1}}{{2\sqrt {{\beta _a}{\beta _b}} }},\Upsilon _{down}^{th}} \right).
\end{align}

By substituting \eqref{B.2} and \eqref{B.3} into \eqref{B.1}, we obtain \eqref{dmp}.

\numberwithin{equation}{section}
\section*{Appendix~C: Proof of Lemma~\ref{lemma2}} \label{Appendix:C}
\renewcommand{\theequation}{C.\arabic{equation}}
\setcounter{equation}{0}

For mmWave scenarios, with the aid of eqs. \eqref{laplaceup} and (26), the Laplace transform of intra-cluster interference in uplink phase \eqref{C.1} can be expressed at the top of next page.
\begin{figure*}[!t]
\normalsize
\begin{align}\label{C.1}
&\mathcal{L}_a\left( s \right) = \mathbb{E}\Big[ {\exp \Big( { - s\sum\limits_{u \in {U_{{v_0}}}\backslash{u_0}} {{\mathrm{L}}\left( {\left\| u \right\|\left| {{h_u},h_{{v_0}}^{up}} \right.} \right){G_u}{{\left| {{{\hat h}_u}} \right|}^2}} } \Big)} \Big]\nonumber \\
&\mathop  = \limits^{(a)} {\mathbb{E}_u}\Big[ {\prod\limits_{u \in {\mathbb{U}_{{v_0}}}\backslash{u_0}} {\sum\limits_{\gamma_1  \in {\mathbb{Z}^*}} {\int_{\frac{{{\gamma _1}}}{{\sqrt {{\beta _a}{\beta _b}} }}}^{\frac{{{\gamma _1} + 1}}{{\sqrt {{\beta _a}{\beta _b}} }}} {{\mathbb{E}_{{{\hat h}_u}}}\Big[ {\exp \left( { - s{\mathrm{L}}\left( {w\left| {{h_u},{h_{v_0}^{up}}} \right.} \right){G_u}{{\left| {{{\hat h}_u}} \right|}^2}} \right)} \Big]f_n^\Omega ( w )dw} } \left| {w = \left\| u \right\|} \right.} } \Big]\nonumber \\
&\mathop  = \limits^{(b)} {\mathbb{E}_u}\Big[ {\prod\limits_{u \in {\mathbb{U}_{{v_0}}}\backslash{u_0}} {\underbrace {\sum\limits_{\gamma_1  \in {\mathbb{Z}^*}} {\int_{\frac{{{\gamma _1}}}{{\sqrt {{\beta _a}{\beta _b}} }}}^{\frac{{{\gamma _1} + 1}}{{\sqrt {{\beta _a}{\beta _b}} }}} {\left( {{\Lambda ^L}\left( {w,s} \right) + {\Lambda ^N}\left( {w,s} \right)} \right)f_n^\Omega \left( w \right)dw} } }_{O_a(s)}} } \Big],
\end{align}
\hrulefill \vspace*{0pt}
\end{figure*}
In \eqref{C.1}, (a) calculates the expectation of communication distance $w$. (b) computes the expectation of the normalized Gamma variable $| {{{\hat h}_u}} |^2$. If the number of users in one cluster is fixed as $\bar N$, the number of interfering devices is $(\bar N-1)$. Due to the independence of each user, \eqref{C.1} can be further derived as
\begin{align}\label{C.2}
\mathcal{L}_a\left( s \right) = {\left( {O_a(s)} \right)^{\bar N - 1}}.
\end{align}

If this number follows a Poisson distribution with the mean $\bar n$, the number of interfering devices, in this case, has the mean $(\bar n-1)$. Therefore, \eqref{C.1} is changed to
\begin{align}\label{C.3}
\mathcal{L}_a\left( s \right) = &\sum\limits_{k = 0}^\infty  {{{\left( { {O_a(s)} } \right)}^k}} \frac{{{{\left( {\bar n - 1} \right)}^k}\exp \left( { - \left( {\bar n - 1} \right)} \right)}}{{k!}}\nonumber \\
=&\exp \left( { - \left( {\bar n - 1} \right)\left( {1 - O_a(s)} \right)} \right)\nonumber \\
&\times \sum\limits_{k = 0}^\infty  {\frac{{{{\left( {O_a(s)\left( {\bar n - 1} \right)} \right)}^k}\exp \left( { - O_a(s)\left( {\bar n - 1} \right)} \right)}}{{k!}}}\nonumber \\
\mathop  = \limits^{(c)} &\exp \left( { - \left( {\bar n - 1} \right)\left( {1 - O_a(s)} \right)} \right),
\end{align}
where (c) calculates the CDF of the Poisson variable with the mean $(O_a(s)(\bar n-1))$. By substituting \eqref{C.2} and \eqref{C.3} into \eqref{C.1}, we obtain \textbf{Lemma~\ref{lemma2}}.

\numberwithin{equation}{section}
\section*{Appendix~D: Proof of Lemma~\ref{lemma3}} \label{Appendix:D}
\renewcommand{\theequation}{D.\arabic{equation}}
\setcounter{equation}{0}

Regarding mmWave scenarios, the Laplace transform of inter-cluster interference can be derived with the aid of \eqref{laplaceup} and (27), which is given by
\begin{align}\label{D.1}
&\mathcal{L}_e\left( s \right)\nonumber\\
= &\mathbb{E}\Big[ {\exp \Big( { - s\sum\limits_{v \in \Phi _v^{up}\backslash{v_0}} {\sum\limits_{u \in {\mathbb{U}_v}} {{\mathrm{L}}\left( {\left\| \left\| u \right\| \right\|\left| {{h_u},h_{{v_0}}^{up}} \right.} \right){G_u}{{\left| {{{\hat h}_u}} \right|}^2}} } } \Big)} \Big]\nonumber \\
\mathop  = \limits^{(a)}& {\mathbb{E}_v}\Big[ {\prod\limits_{v \in \Phi _v^{up}\backslash{v_0}} {{\mathbb{E}_u}\Big[ {\prod\limits_{u \in {\mathbb{U}_v}} }}}\nonumber\\
&\times {{{{\underbrace {\sum\limits_{{\gamma _1} \in {\mathbb{Z}^*}} {\int_{\frac{{{\gamma _1}}}{{\sqrt {{\beta _a}{\beta _b}} }}}^{\frac{{{\gamma _1} + 1}}{{\sqrt {{\beta _a}{\beta _b}} }}} {\left( {{\Lambda ^L}\left( {g,s} \right) + {\Lambda ^N}\left( {g,s} \right)} \right)f_Q^\Omega \left( {g\left| q \right.} \right)dg} } }_{ O_e(s,q)}} } \Big]} } \Big],
\end{align}
where $g = \left\| {u + v} \right\|$ and $q = \left\| v \right\|$. (a) follows the similar procedure of (a) and (b) in Appendix C. The difference is the distance distributions in this case obey \eqref{Thoe} and \eqref{Mate}. If the number of users in each cluster is fixed as $N$, the number of interfering devices in every other cluster is same with $\bar N$. Therefore the Laplace transform of inter-cluster interference can be expressed as
\begin{align}\label{D.2}
&\mathcal{L}_e\left( s \right) = {\mathbb{E}_v}\big[ {\prod\limits_{v \in \Phi _v^{up}\backslash{v_0}} {{{\left( {O_e(s,q)} \right)}^{\bar{N}}}} } \big]\nonumber \\
&\mathop  = \limits^{(b)} \exp \left( { - 2\pi \lambda _v^{up}\int_0^\infty  {\left( {1 - {{\left( {\hat O_e(s,q)} \right)}^{\bar{N}}}} \right)} qdq} \right),
\end{align}
where (b) follows the probability generating function of PPP~\cite{stoyanstochastic,8016632}.

In terms of the Poisson distributed number of users in one cluster, the average number of interfering users in each cluster is $\bar n$. \eqref{D.1} can be further deduced as
\begin{align}\label{D.3}
&\mathcal{L}_e\left( s \right)\nonumber\\
\mathop  = \limits^{(b)} &\exp\Big({ { - 2\pi \lambda _v^{up}\int_0^\infty  {\left( {1 - \exp \left( { - \bar n\left( {1 - \hat O_e(s,q)} \right)} \right)} \right)} qdq} }\Big).
\end{align}

By substituting \eqref{D.2} and \eqref{D.3} into \eqref{D.1}, we obtain \textbf{Lemma~\ref{lemma3}}.

\bibliographystyle{IEEEtran}
\bibliography{mybib}
\end{document}